\documentclass[12pt,preprint]{aastex}
\usepackage{rotating}

\newcommand{\kms}{\mbox{km\,s$^{-1}$}}
\newcommand{\Kkms}{\mbox{K\,km\,s$^{-1}$}}
\newcommand{\MSX}{{\it {MSX}}}
\newcommand{\Spitzer}{{\it {Spitzer}}}
\newcommand{\Msun}{$M_{\odot}$}

\newcommand{\hi}{\mbox{$\mathrm{H\,{\scriptstyle {I}}}$}}

\newcommand{\co}{\mbox{$^{12}${\rmfamily CO}{~$J=1\rightarrow 0$}}}

\newcommand{\cont}{\mbox{$^{12}${\rmfamily CO}}}
\newcommand{\tcont}{\mbox{$^{13}${\rmfamily CO}}}

\newcommand{\tco}{\mbox{$^{13}${\rmfamily CO}{~$J=1\rightarrow 0$}}}
\newcommand{\vlsr}{V$_{LSR}$}

\newcommand{\tastar}{\mbox{T$_{A}^{*}$}}
\newcommand{\tmb}{\mbox{T$_{mb}$}}
\newcommand{\tex}{\mbox{T$_{ex}$}}
\newcommand{\Tk}{\mbox{T$_{k}$}}

\newcommand{\degree}{$^{\circ}$}
\newcommand{\cms}{${\rm cm}^{-2}$}
\newcommand{\cmc}{${\rm cm}^{-3}$}
\newcommand{\hh}{H$_{2}$}
\newcommand{\bufcrao}{Boston University--Five College Radio Astronomy Observatory}
\newcommand{\grs}{Galactic Ring Survey}
\newcommand{\lb}{{\mbox{($\ell, b$)}}}
\newcommand{\lv}{{\mbox{($\ell, v$)}}}

\slugcomment{}

\shorttitle{GRS Molecular Clouds}
\shortauthors{Rathborne et al.}

\begin{document}

\title{Molecular clouds and clumps in the Boston University--Five College Radio Astronomy Observatory Galactic Ring Survey}
\author{J. M. Rathborne\footnote{Current address: Harvard-Smithsonian Center for Astrophysics, Mail Stop 42, 60 Garden Street, Cambridge, MA 02138, USA}, A. M. Johnson, J. M. Jackson, R. Y. Shah}
\affil{Institute for Astrophysical Research, Boston University, Boston, MA 02215; rathborn@bu.edu, alexj@bu.edu, jackson@bu.edu, ronak@bu.edu}
\and
\author{R. Simon}
\affil{I.Physikalisches Institut, Universit\"at zu K\"oln, 50937 K\"oln, Germany; simonr@ph1.uni-koeln.de} 

\begin{abstract}
The \bufcrao\, (BU-FCRAO) \grs\, (GRS) of \tco\, emission covers
Galactic longitudes 18$\arcdeg < \ell < 55.7\arcdeg$ and Galactic
latitudes $|b| \leq 1\arcdeg$. Using the SEQUOIA array on the FCRAO
14m telescope, the GRS fully sampled the \tcont\, Galactic emission
(46\arcsec\, angular resolution on a 22\arcsec\, grid) and achieved a
spectral resolution of 0.21\,\kms. Because the GRS uses \tcont, an
optically thin tracer, rather than \cont, an optically thick tracer,
the GRS allows a much better determination of column density and also
a cleaner separation of velocity components along a line of
sight. With this homogeneous, fully-sampled survey of \tcont\,
emission, we have identified 829 molecular clouds and 6124 clumps
throughout the inner Galaxy using the CLUMPFIND algorithm.  Here we
present details of the catalog and a preliminary analysis of the
properties of the molecular clouds and their clumps. Moreover, we
compare clouds inside and outside of the 5~kpc ring and find that
clouds within the ring typically have warmer temperatures, higher
column densities, larger areas, and more clumps compared to clouds
located outside the ring. This is expected if these clouds are
actively forming stars. This catalog provides a useful tool for the
study of molecular clouds and their embedded young stellar objects.
\end{abstract}
\keywords{catalogs--molecular data--ISM:clouds--ISM: molecules--Galaxy: kinematics and dynamics}
\section{Introduction}

The \bufcrao\, (BU--FCRAO) \grs\, (GRS) is a survey of Galactic
\tco\, emission toward the first Galactic quadrant  \citep{Jackson06}. 
One of the primary goals of the GRS was the study the Milky Way's
so-called 5\,kpc molecular ring.  Discovered as a peak in the radial
Galactic CO distribution midway between the sun and the Galactic
center \citep{Burton76,Scoville75}, the 5~kpc molecular ring contains
$\sim$ 70\%\, of the molecular material within the solar circle (M
$\sim$ 2 $\times$ 10$^{9}$\,\Msun; \citealp{Combes91}). It is the
prominent feature in molecular maps of the Galaxy and is the location
of the majority of Galactic star formation
\citep{Burton76,Robinson84,Clemens88}.

Because the 5\,kpc ring covers such a large region of sky, a large,
high angular resolution, sensitive survey is required to reveal the
molecular clouds and dense clumps where star formation occurs. In
order to be able to study their properties, the BU--FCRAO GRS was
designed to map the \tcont\, molecular line emission over a large
portion of the 5 kpc ring and, in doing so, answer a number of
fundamental questions that remain about the 5 kpc ring.  For instance,
how many molecular clouds does the 5 kpc ring contain?  What are their
properties?  Do molecular clouds in the star-forming ring differ from
clouds elsewhere in the Galaxy?

The GRS provides a unique database to identify star-forming molecular
clouds and their dense clumps.  The first step in such studies is to
isolate, identify, and catalog the molecular clouds and their
clumps. Combined with distance determinations, this catalog can
characterize the cloud masses, sizes, line widths, and densities in a
range of Galactic environments.  This paper describes a catalog of
molecular clouds and clumps identified using \tcont\, data from the
GRS which provides a new, homogeneous database for the study of the
large- and small-scale structure of molecular gas throughout the inner
Milky Way (e.g.,\
\citealp{Simon01}). The complete catalog is available in electronic
form\footnote{see
http://www.bu.edu/galacticring/molecular\_clouds.html}. In total, we
identify 829 clouds and 6124 clumps.  A preliminary analysis of their
general characteristics is also presented. In particular, we compare
the properties of those clouds located inside the 5 kpc molecular ring
to those outside the ring.

\section{The BU--FCRAO Galactic Ring Survey}

The BU--FCRAO GRS \citep{Jackson06} mapped Galactic \tco\, emission
using the SEQUOIA multi-pixel array on the FCRAO 14 m telescope.  The
survey extended in Galactic longitude from $\ell =$ 18$\arcdeg$ to
55.7$\arcdeg$ and in Galactic latitude from $|b| \leq 1\arcdeg$, for a
total of 75.4 square degrees. The survey's velocity coverage was $-$5
to 135\,\kms\, for Galactic longitudes $\ell \leq 40\arcdeg$ and $-$5
to 85\,\kms\, for Galactic longitudes $\ell > 40\arcdeg$. The typical
rms sensitivity was $\sigma$(\tastar) $\sim$ 0.13\,K.  The survey
comprises a total of 1,993,522 spectra.  The data are available to the
community at www.bu.edu/galacticring or in DVD form by request.

Unlike most previous surveys of the inner Galaxy, the GRS is fully
sampled (46\arcsec\, angular resolution on a 22\arcsec\, grid), has a
high spectral resolution (0.21\,\kms), and traces \tco\, emission.
Compared to \cont, the \tcont\, molecule is $\sim$ 50 times less
abundant and, thus, has a much lower optical depth. As a result,
\tcont\, is a much better tracer of column density. Moreover, because
it suffers less from line blending and self-absorption, it can
separate clouds along the same line of sight more easily than
\cont. Because of these improvements over previous surveys the GRS can
detect many new structures and cloud cores previously missed by older
\cont\, surveys (e.g.,\ \citealp{Sanders86,Dame01}).

For more details of the telescope and instrumental parameters, the
observing modes, the data reduction processes, and the emission and
noise characteristics of the dataset see \cite{Jackson06}.

\section{Identifying molecular clouds and clumps}

We use the automated algorithm CLUMPFIND \citep{Williams94} to
identify molecular clouds and clumps within the GRS.  CLUMPFIND
searches through a three-dimensional ($\ell$, $b$, $v$) data cube
using iso-brightness surfaces to identify contiguous emission features
without assuming an a priori shape. In this respect, it is more
flexible than other cloud identification algorithms that artifically
decompose the emission features into three-dimensional Gaussian profiles
(e.g., GAUSSCLUMPS).

CLUMPFIND begins the search for clouds at the voxel with the peak
brightness in the data cube. The algorithm steps down from this peak
brightness in levels referred to as `contour increments'. During the
first iteration the algorithm finds all contiguous voxels with
brightnesses between the peak value and the next level (defined as the
peak value minus the contour increment).  If these contiguous voxels
are isolated they are assigned to a new cloud. If they are contiguous
with a previously identified cloud then they are assigned to that
pre-existing cloud. The algorithm iterates until it reaches a minimum
brightness level. At this lowest level, CLUMPFIND does not search for
new clouds, but instead adds all emission at that level into
pre-existing clouds. This procedure is straightforward for isolated
emission peaks, but complicated for blended emission features. For
blended emission, features are separated based on a
`friends-of-friends' algorithm (see
\citealp{Williams94} for more details).

For each cloud (or clump) that is identified, CLUMPFIND gives the
position of the peak of the emission for the cloud in ($\ell$, $b$,
$v$), the peak temperature, the FWHM extent of the cloud in ($\ell$,
$b$, $v$), an estimate of the radius of the cloud (assuming the total
number of voxels form a sphere), the total number of voxels, and the
sum of the emission within the voxels.  Moreover, a three-dimensional
($\ell$, $b$, $v$) data cube is also created which is identical to the
input data cube in ($\ell$, $b$, $v$) space. In this output cube,
however, the temperature scale in each voxel is replaced by an integer
which corresponds to the cloud to which it is assigned. This output
cube is extremely useful for identifying the exact voxels associated
with each individual cloud and is used extensively in the analysis
described below.

\subsection{Clouds}

Giant molecular clouds (GMCs) are generally defined as extended
molecular line emission features with typical properties as outlined
in Table~\ref{properties}. Molecular clouds do not have uniform
densities, but are observed to be fragmented on all size scales. We
refer to the dense regions within GMCs where clusters may form as
`clumps' and the smaller, denser sites where the individual star
formation occurs as `cores'. Table~\ref{properties} compares the
typical properties of each of these structures.

\subsubsection{Smoothing of the GRS data}

\begin{figure}
\includegraphics[angle=-90,width=0.7\textwidth]{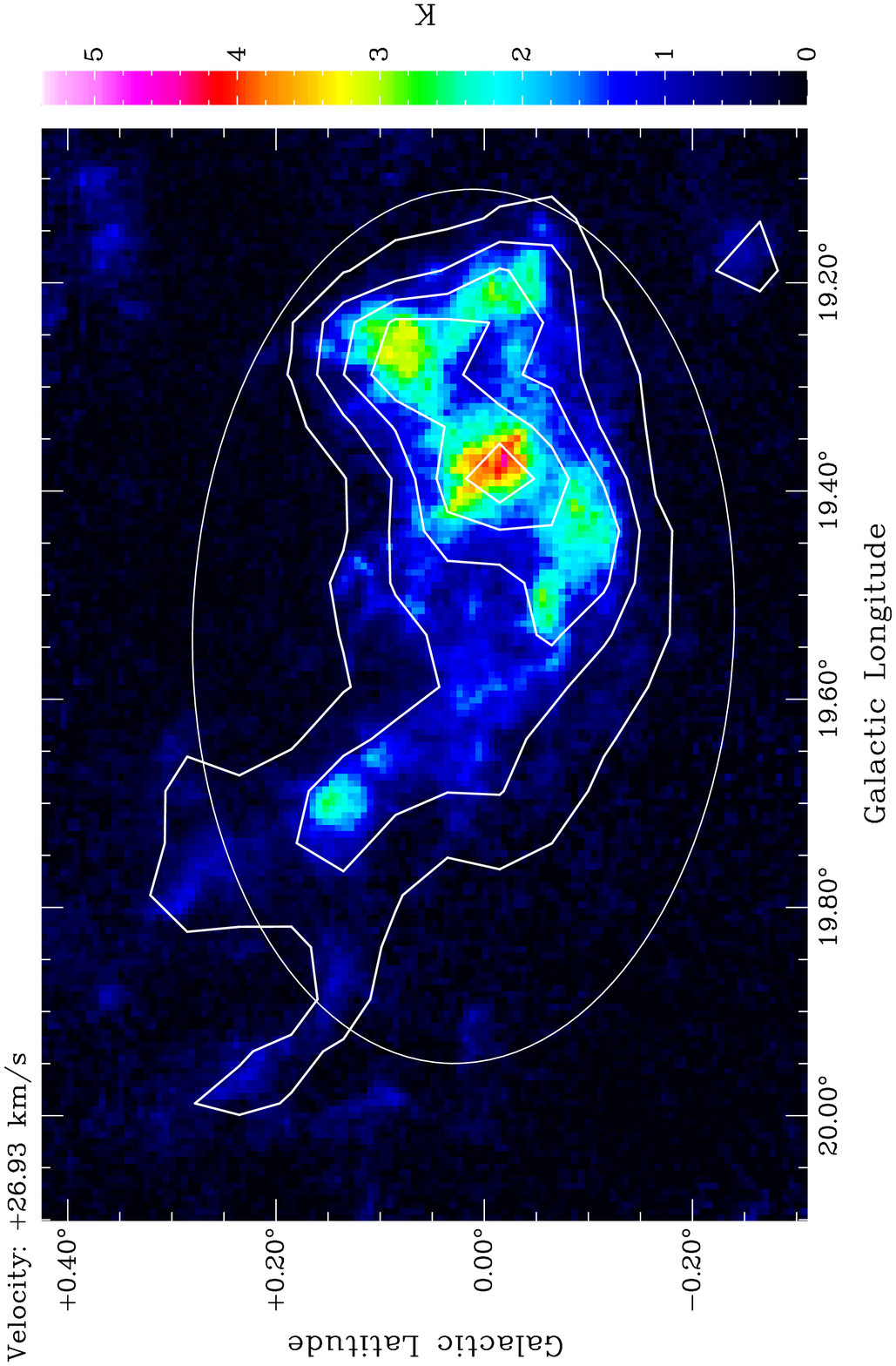}\\
\includegraphics[angle=-90,width=0.7\textwidth]{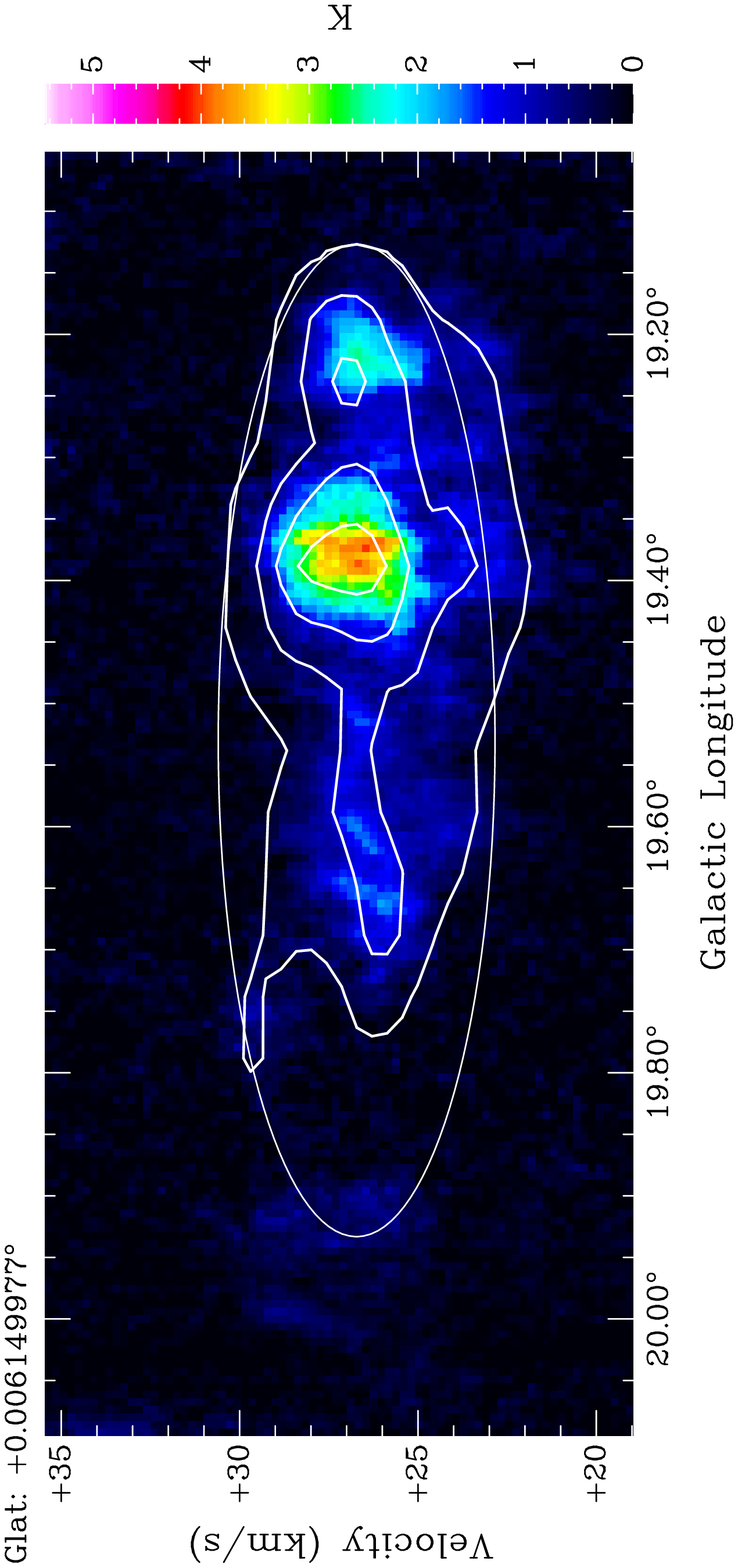}\\
\caption{\label{grs-smooth} An example of the comparison between the full angular 
    and spectral resolution GRS data (in color scale) and the smoothed
    data (as contours) for GRSMC G019.39$-$00.01. The top panel shows a
    single \lb\, channel while the lower panel shows a single plane in
    \lv\, space. The smoothed data were generated by smoothing both
    spatially (6\arcmin\, on a 3\arcmin\, grid) and spectrally
    (0.6\,\kms) the original data. The smoothed data have an rms
    sensitivity of $\sigma$(\tastar) $\sim$ 0.01\,K.  The contour
    levels are 0.2 to 3.0 K in steps of 0.3 K on the \lb\, image and
    0.2 to 2.0 in steps of 0.4 K in the \lv\, image. The white
    ellipses mark the approximate extent of the cloud in \lb\, and
    \lv\, space.}
\end{figure}

To improve the signal-to-noise and to better identify faint, extended
molecular clouds we smoothed the GRS data both spatially (to a
resolution of 6\arcmin) and spectrally (to a resolution of 0.6\,\kms)
using Gaussian kernels. Because this smoothing was moderate, it did not
cause a significant loss of information in $\ell$, $b$, or $v$. In
particular, this smoothing did not cause molecular clouds to be
artifically blended along the line-of-sight.  The smoothing improved
the rms sensitivity from $\sim$0.13\,K for the original data to $\sim$
0.01\,K for the smoothed data. The smoothed data were then re-sampled
on a 3\arcmin\, grid. Figure~\ref{grs-smooth} shows a comparison
between the original and smoothed data in both \lb\, and \lv\, space.

\subsubsection{Modification of the CLUMPFIND algorithm}

The CLUMPFIND algorithm was originally designed to identify the clumpy
substructure within individual molecular clouds. To run CLUMPFIND, the
user must specify two input parameters; the contour increment and the
lowest brightness level.  In the original algorithm, the lowest
brightness level is hardwired to be a multiple of the contour
increments which are forced to be evenly spaced, from the peak
brightness to the minimum level. Because of the large dynamic range of
the GRS data, however, the algorithm in this original form was unable
to identify simultaneously both extended, diffuse clouds as well as
compact, bright clouds. Moreover, it could not easily separate clouds
that appeared distinct, especially at the lowest emission levels.  For
instance, by specifying a large contour increment (e.g.\ 0.7 K)
CLUMPFIND would identify the brightest clouds in the GRS as one
contiguous feature, but missed or merged many of the fainter, more
extended clouds. However, with a smaller contour increment (e.g.\ 0.1
K), CLUMPFIND would better identify and separate the fainter, extended
clouds, but the largest, brightest clouds would be artificially
dissected and identified as numerous separate clouds.

Thus, in order to identify molecular clouds within the GRS at all
emission levels, we modified the CLUMPFIND algorithm. For this purpose
we required a large contour increment, but a small minimum brightness
level. To achieve this, we changed the minimum brightness level to be
a user input variable independent of the contour increment. In this
scheme, we retained evenly spaced contour levels between the peak and
minimum brightness levels as per the original algorithm but could
additionally search for clouds to a lower brightness level. This
scheme allowed us to identify both faint, extended clouds and bright,
compact clouds with the same input parameters.

\subsubsection{Selection of CLUMPFIND input parameters}
\label{iding-clouds}

With this modified version of CLUMPFIND, we tested a range of contour
increments and lowest brightness levels to determine which set of
parameters would best identify molecular clouds.  Because of the
inherent difficulties in determining where one cloud ends and the next
begins, different choices of these parameters will certainly result in
different output catalogs. However, regardless of the exact
parameters, the bright, large molecular clouds are typically always
recovered; the differences in the catalogs is most noticeable in the
separation of nearby emission features and for the faint, small
clouds.

To test the performance of the algorithm we completed a series of
simple tests to compare the clouds identified with CLUMPFIND to the
original dataset. In essence we replaced the CLUMPFIND clouds with
elliptical three-dimensional Gaussian models with the same peak
temperature, center position, size, orientation, line width and voxels
of each cloud identified by CLUMPFIND. We then subtracted these
Gaussian models from the original dataset and examined the
residuals. We describe this process below.

We ran three representative regions within the smoothed GRS (each covering
$\Delta \ell \times \Delta b$ = 2\degree$\times$2\degree) through
CLUMPFIND, varying both the contour increment and the minimum
brightness level. The test regions were selected to span a range from
crowded and complex to sparse and simple emission features. We varied
the contour increment from 0.1 (10$\sigma$) to 0.7 K in steps of 0.1 K
and the minimum brightness level from 0.05 (5$\sigma$) to 0.4 K in
steps of 0.05~K and examined the output from all combinations of these
two parameters.

Comparisons between the Gaussian model and the original dataset reveal
that, for simple unblended emission features, the Gaussian model
matched the data very well: the mean residual between the data and the
model was $\sim$ 2$\sigma$ of the data. Not surprisingly, the mean
residuals were smallest when the minimum values of the contour
increment and lowest minimum brightness level were used (0.10 K and
0.05 K respectively). Put another way, the Gaussian model reproduced
the data best when a larger number of clouds were found (e.g., a
residual $\sim$0.023 K for N$_{\rm{clouds}}$ = 115 compared to a
residual $\sim$0.028 K for N$_{\rm {clouds}}$ = 9).

In regions of bright, complicated emission features and blended lines,
the mean residual between the original dataset and the Gaussian model
were slightly higher (residual$\sim$ 0.065 K). For these regions, the
mean residuals were insensitive to the choice of CLUMPFIND input
parameters.  For the range of CLUMPFIND input parameters tested, the
residuals varied by only $\sim$ 0.01 K (1$\sigma$).  Thus, for
identifying the brightest molecular clouds in the dataset
($>10\sigma$) we found the exact choice of parameters is not crucial.

\begin{figure}
\includegraphics[angle=-90,width=0.5\textwidth]{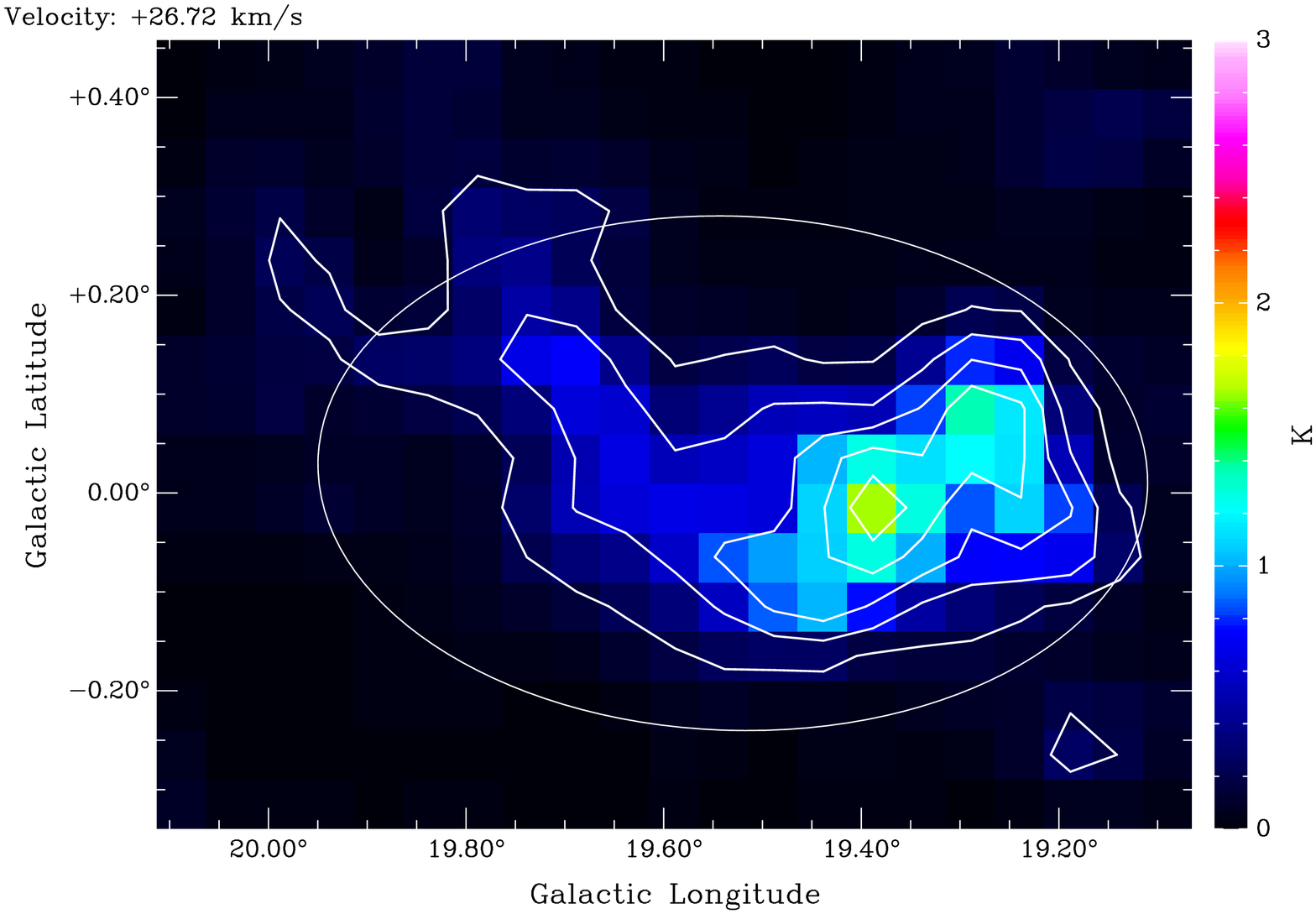}
\includegraphics[angle=-90,totalheight=0.25\textheight]{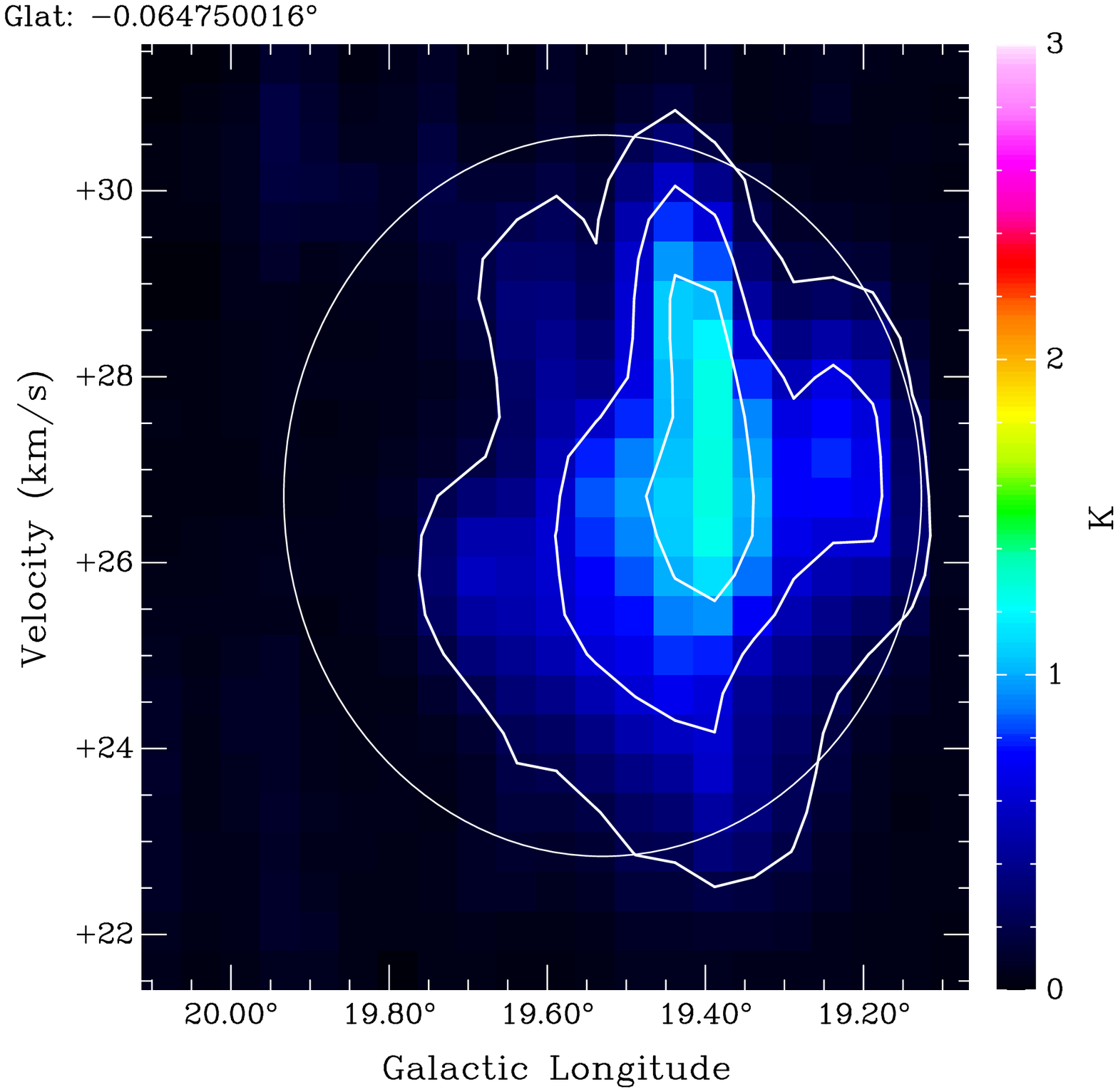}\\
\includegraphics[angle=-90,width=0.5\textwidth]{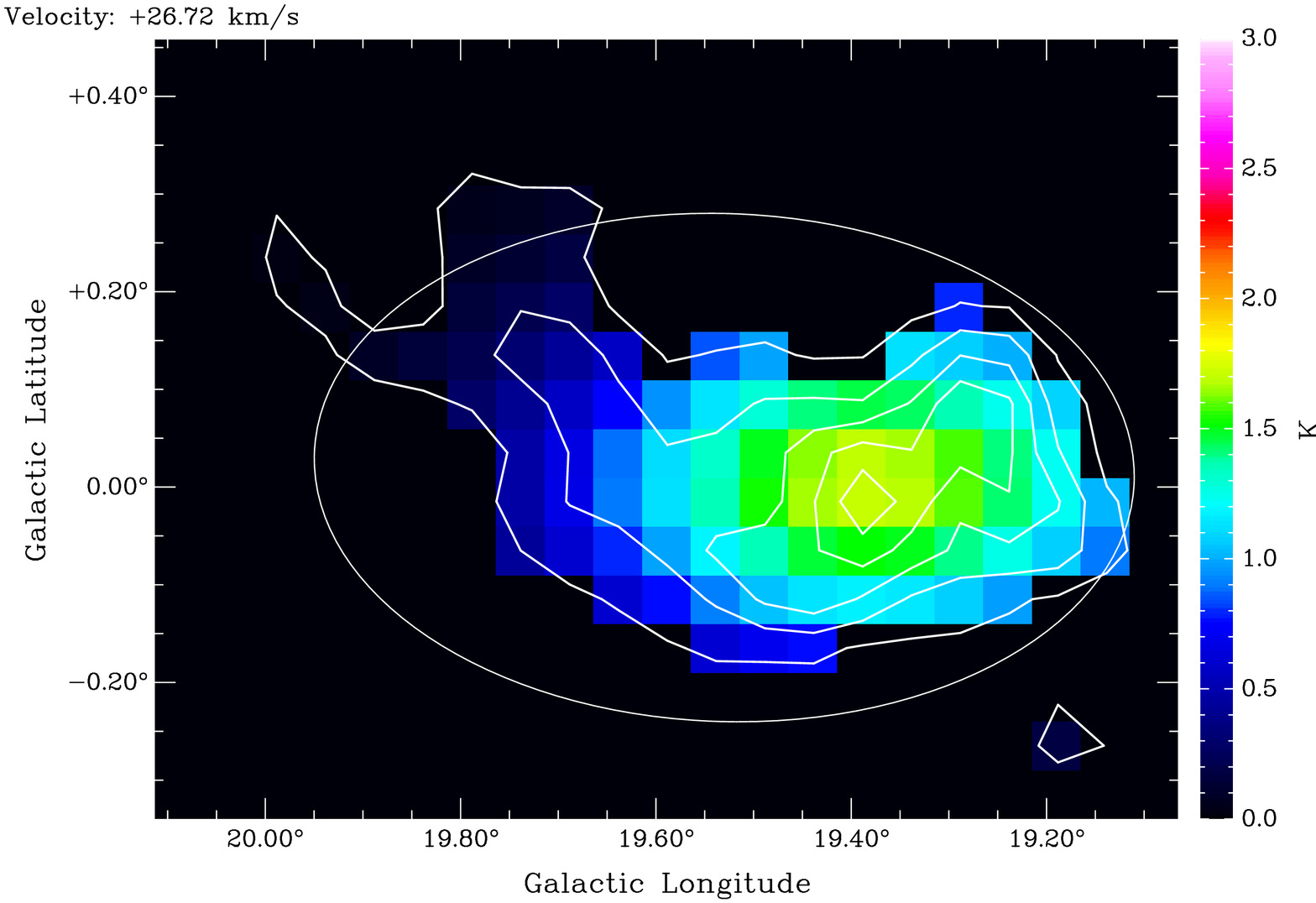}
\includegraphics[angle=-90,totalheight=0.25\textheight]{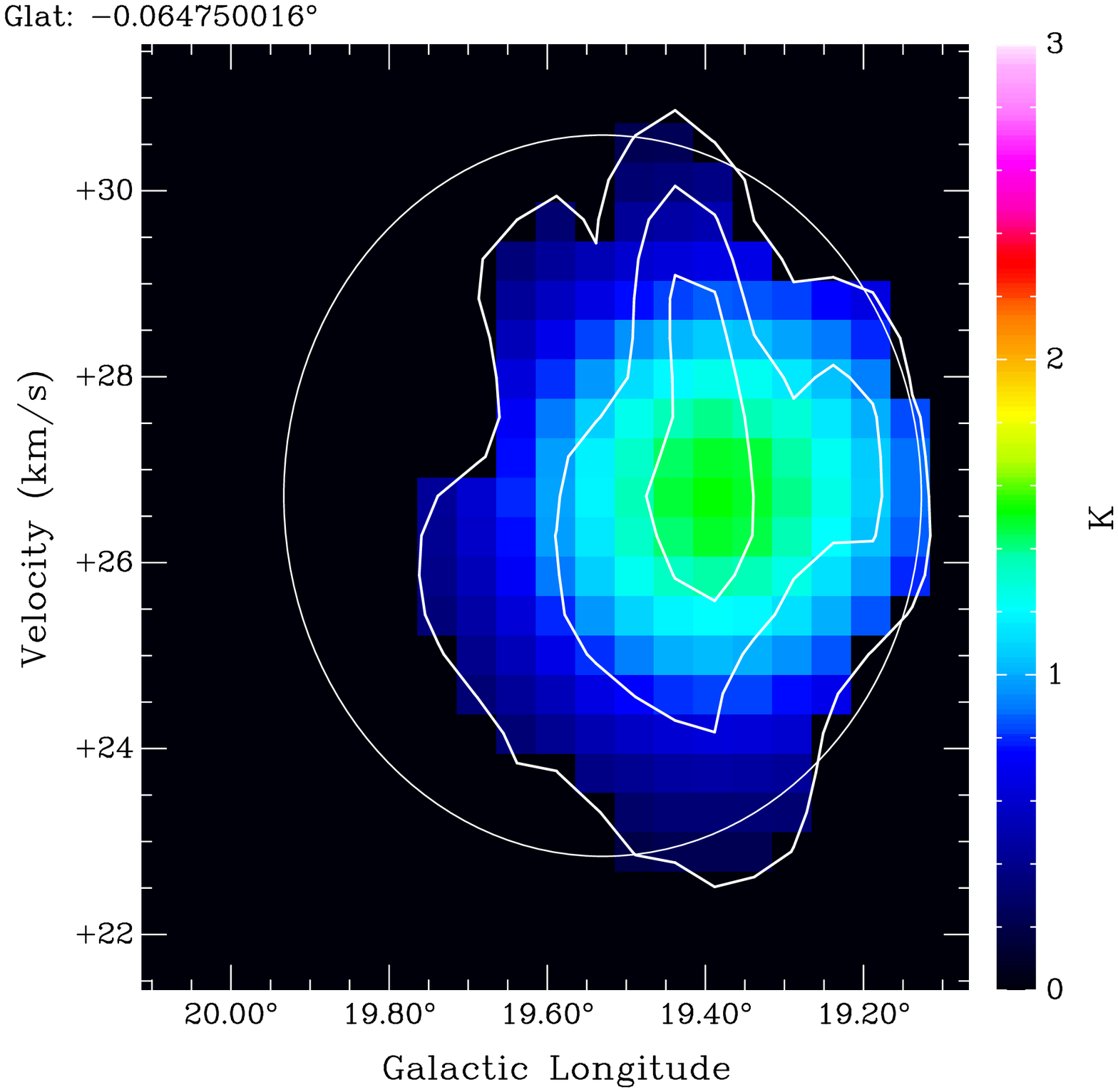}\\
\includegraphics[angle=-90,width=0.5\textwidth]{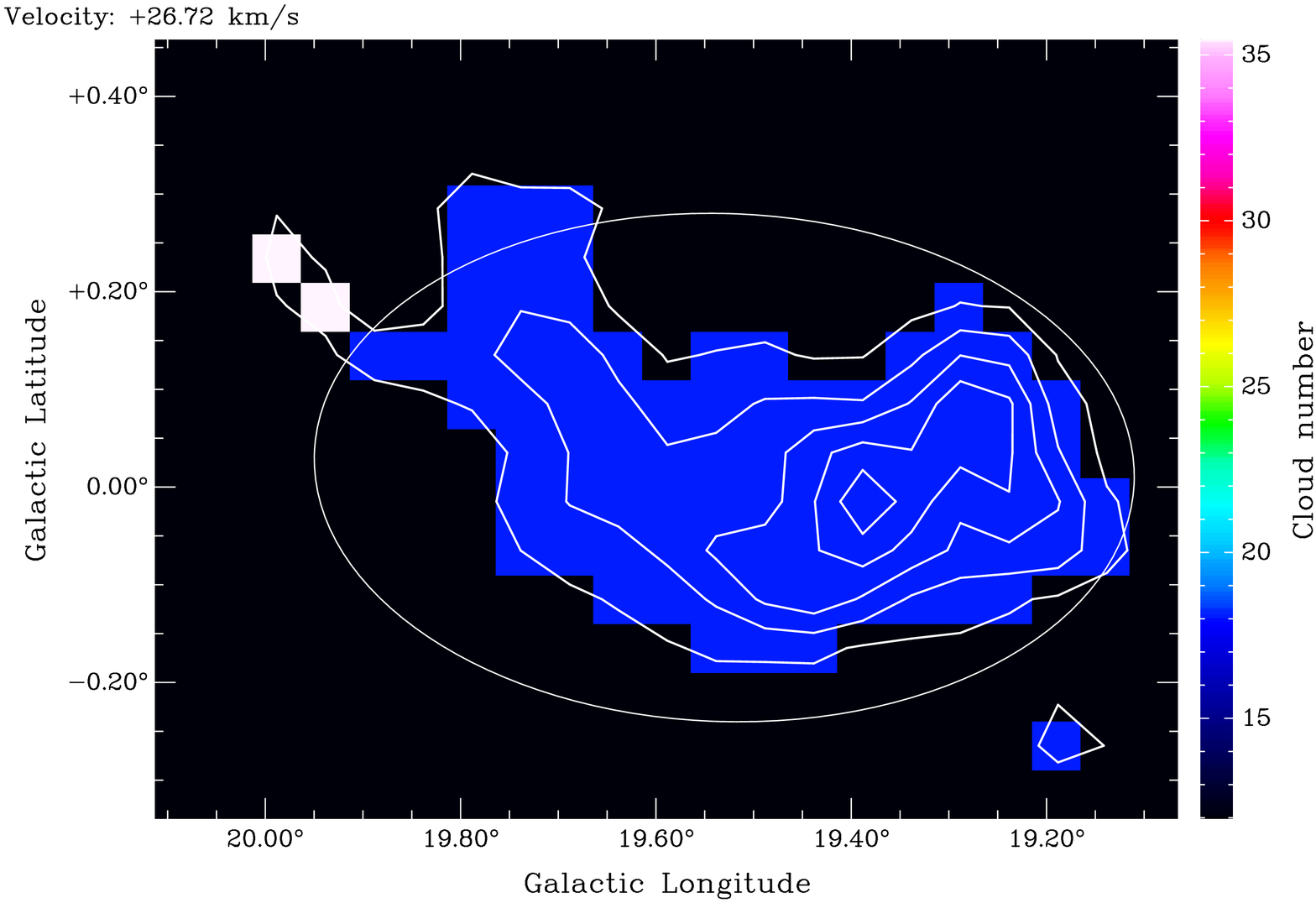}
\includegraphics[angle=-90,totalheight=0.25\textheight]{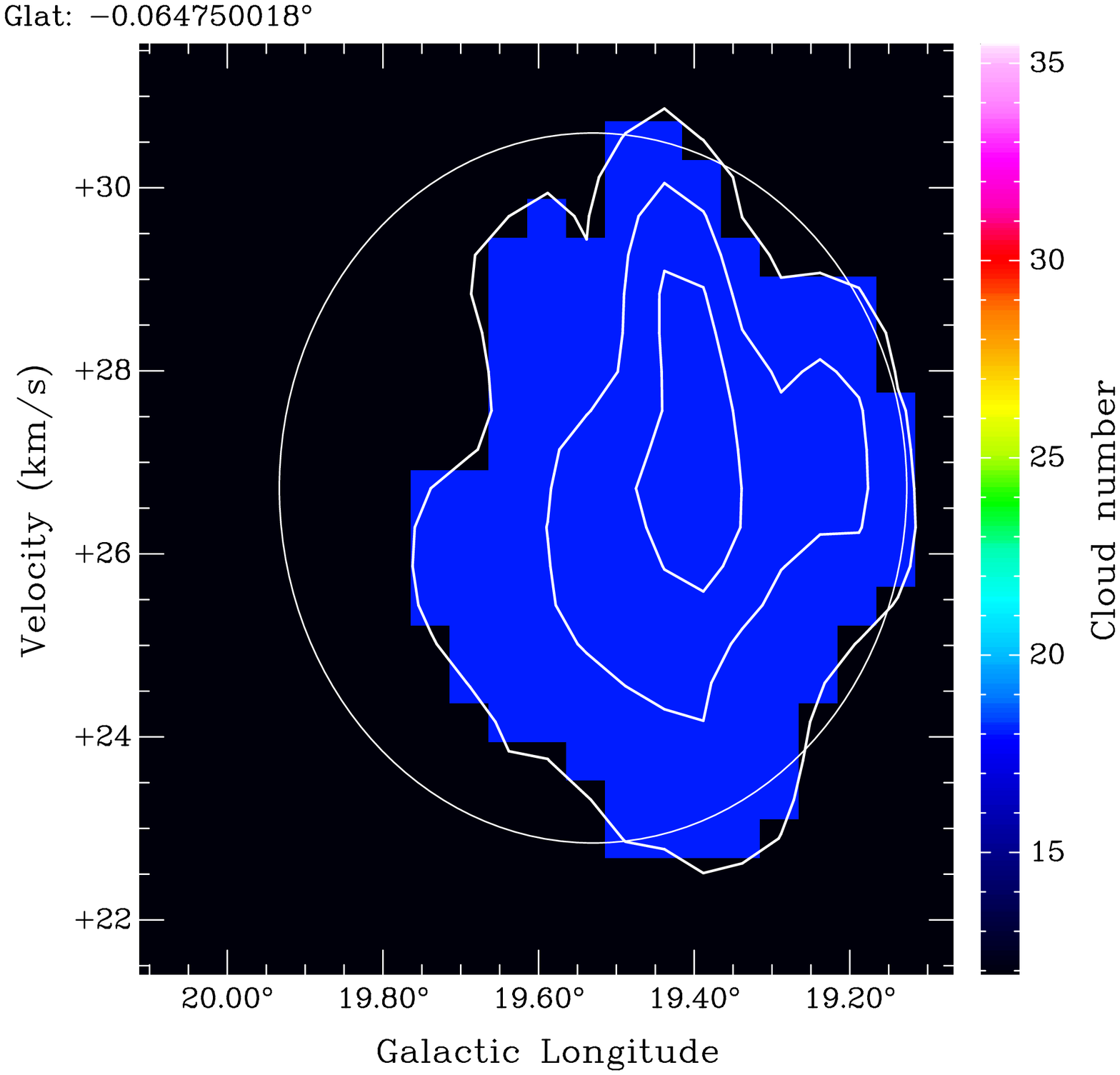}\\
\caption{\label{clfind-eg} An example of the identification of an
   isolated molecular cloud in ($\ell$, $b$) and ($\ell$, $v$)
   space. The left panels show single velocity channel in the ($\ell$,
   $b$) image, while the right panels show a single Galactic latitude
   plane in the ($\ell$, $v$) image. The top panels show the smoothed
   GRS data in color scale. On all images the contours are the
   smoothed data, while the white ellipses mark the approximate extent
   of the cloud as determined by CLUMPFIND; the major and minor axes
   are equivalent to the projected extent of the cloud in each
   direction. The middle panels show the corresponding Gaussian model,
   which was generated using the peak temperature, center position,
   size, orientation, line width, and voxels output from
   CLUMPFIND. The lower panels show the CLUMPFIND output cube; the
   color scale in these images represent the voxels that are assigned
   to each cloud for each particular channel or plane.}
\end{figure}

Figure~\ref{clfind-eg} shows an example of the identification of an
isolated molecular cloud in ($\ell$, $b$) and ($\ell$, $v$)
space. Because of the difficulties of displaying three-dimension data
cubes in a two-dimensional image, we have selected to show a single
velocity channel in the ($\ell$, $b$) image (left panels) and a single
Galactic latitude plane in the ($\ell$, $v$) image (right panels). The
top panels show an isolated molecular cloud in the smoothed GRS data
(in color scale and as contours in this and all subsequent
images). The white ellipses mark the approximate extent of the cloud
as determined by CLUMPFIND; the major and minor axes are equivalent to
the projected extent of the cloud in each direction. The middle panels
of Figure~\ref{clfind-eg} show the corresponding Gaussian model, which
was generated using the peak temperature, center position, size,
orientation, and line width output from CLUMPFIND. The lower panels
show the CLUMPFIND output cube; the color scale in these images
represent the voxels that are assigned to each cloud for each
particular channel or plane.

\cite{Williams94} found that CLUMPFIND most accurately
represents the data when the contour increment is set to twice the
1~$\sigma$ noise of the data. Because we are interested in selecting
the largest and brightest molecular clouds in our dataset, we use a
contour increment of twice the 10$\sigma$ noise (rather than
1$\sigma$). Thus, to identify clouds we use a contour increment of
twice the 10$\sigma$ noise in the smoothed data (0.20~K) and a lowest
brightness level of 20$\sigma$ (0.20 K).  Our comparisons of the
Gaussian models with the original data indicated that these parameters
did not produce extremes in either the number of detected clouds or
mean residuals. Thus, we consider them suitable for isolating and
identifying molecular clouds.

The output from CLUMPFIND was also checked by eye against the original
dataset to verify that these parameters could simultaneously identify
and separate both bright and compact molecular clouds in addition to
diffuse and extended molecular clouds. This choice of input parameters
was confirmed to be satisfactory and were adopted to produce the final
catalog. Thus, using a contour increment of 0.20 K and a lowest
brightness level of 0.20 K CLUMPFIND, identified 848 molecular clouds
in the GRS dataset.

For clouds to be selected as real features, CLUMPFIND requires
each cloud to have a minimum number of voxels (we set this number to
16). To be confident we are identifying real molecular clouds, we
imposed additional size criteria on each cloud before inclusion in the
final list. We exclude all clouds that have their measured sizes in
any axis less than or equal to the resolution of the smoothed data in
that axis (i.e., $\Delta \ell$ or $\Delta b$ $\le$ 6\arcmin\, or
$\Delta$~V $\le$ 0.6\,\kms).  We find 18 clouds that meet these
criteria. Thus, when we exclude these, the final catalog contains 829
molecular clouds. In the sections below we characterize the properties
of these clouds.

\subsubsection{Determining sizes of the clouds}
\label{ellipse-fitting}

Because molecular clouds have complicated morphologies, it is
difficult to quantify their shapes and measure their sizes. Although
CLUMPFIND determines the extent of each cloud in Galactic longitude
and latitude, this determination fails to account for the cloud's
orientation. To more accurately describe their shapes, we modeled the
two-dimensional integrated emission for each cloud using a
two-dimensional ellipse fitting routine\footnote{written in IDL by
D. Fanning.}. Two-dimensional integrated intensity maps of each cloud
were produced by summing in velocity over all the voxels identified by
CLUMPFIND for a given cloud. The center position of the cloud was then
determined from the centroid of this velocity integrated image (with
all pixels weighted equally regardless of their integrated
intensity). The semimajor and semiminor axes were determined by
fitting an ellipse to this integrated intensity image. The listed
position angle was measured counter-clockwise from the positive
Galactic longitude axis.  With this convention, a cloud whose major
axis lines along the Galactic plane will have a position angle of
0\degree\, while one lying perpendicular to the Galactic plane will
have a position angle of 90\degree. This method provides a simple
description of the cloud's position, size, and orientation with
respect to the Galactic plane.

We have also calculated the projected two-dimensional area for each
cloud which was determined by multiplying the total number of pixels
within the integrated intensity image by the area of a pixel in square
degrees.

\subsection{Clumps}

Because molecular clouds show structure on all size scales, they
typically contain several compact, dense clumps.  These clumps
presumably represent the compact, dense regions within the molecular
clouds where star and cluster formation may take place.  To identify
these star-forming clumps, we searched for three-dimensional
substructure within each of the 829 molecular clouds in ($\ell$, $b$,
$v$) space. To achieve this we used the full angular and spectral
resolution GRS data and the original CLUMPFIND algorithm.

For each molecular cloud, we search for clumps using only the voxels
for that cloud identified by CLUMPFIND as described in the previous
section.  We also restricted the size of each clump to at least 50
voxels (in total in ($\ell$, $b$, $v$) space) in order to prevent
false detections.

The same technique using three-dimensional Gaussian models (as
described in $\S$\ref{iding-clouds}) was used to determine the best
CLUMPFIND input parameters to identify and separate the clumps within
each of the clouds. We selected 10 molecular clouds to test the
CLUMPFIND input parameters for clump identification. These molecular
clouds were selected to include a range in sizes, emission levels, and
substructures. To test the clump identification, we varied the contour
increment using values of 0.2, 0.26, 0.36, 0.39 K and the lowest
brightness level from 0.52 K to 1.3 K in steps of 1.3 K and examined, by
eye, the output from all combinations of these parameters.

We found a contour increment of 0.26 K (2$\sigma$) and a minimum
brightness level of 1.30 K (10$\sigma$) was best at isolating and
identifying the individual clumps within the clouds. Using these
parameters, we identify, in total, 6135 clumps within these
clouds. Again, for the final catalog, we exclude all clumps that have
sizes smaller than the resolution in any axis (i.e. $\Delta \ell$ or
$\Delta b$ $\le$ 0\fdg01 or $\Delta$ V $\le$ 0.2\,\kms).  Thus, the
final clump catalog contains 6124 entries.


\subsection{Examples of clouds and clumps}

\begin{figure}
\includegraphics[angle=-90,width=0.42\textwidth]{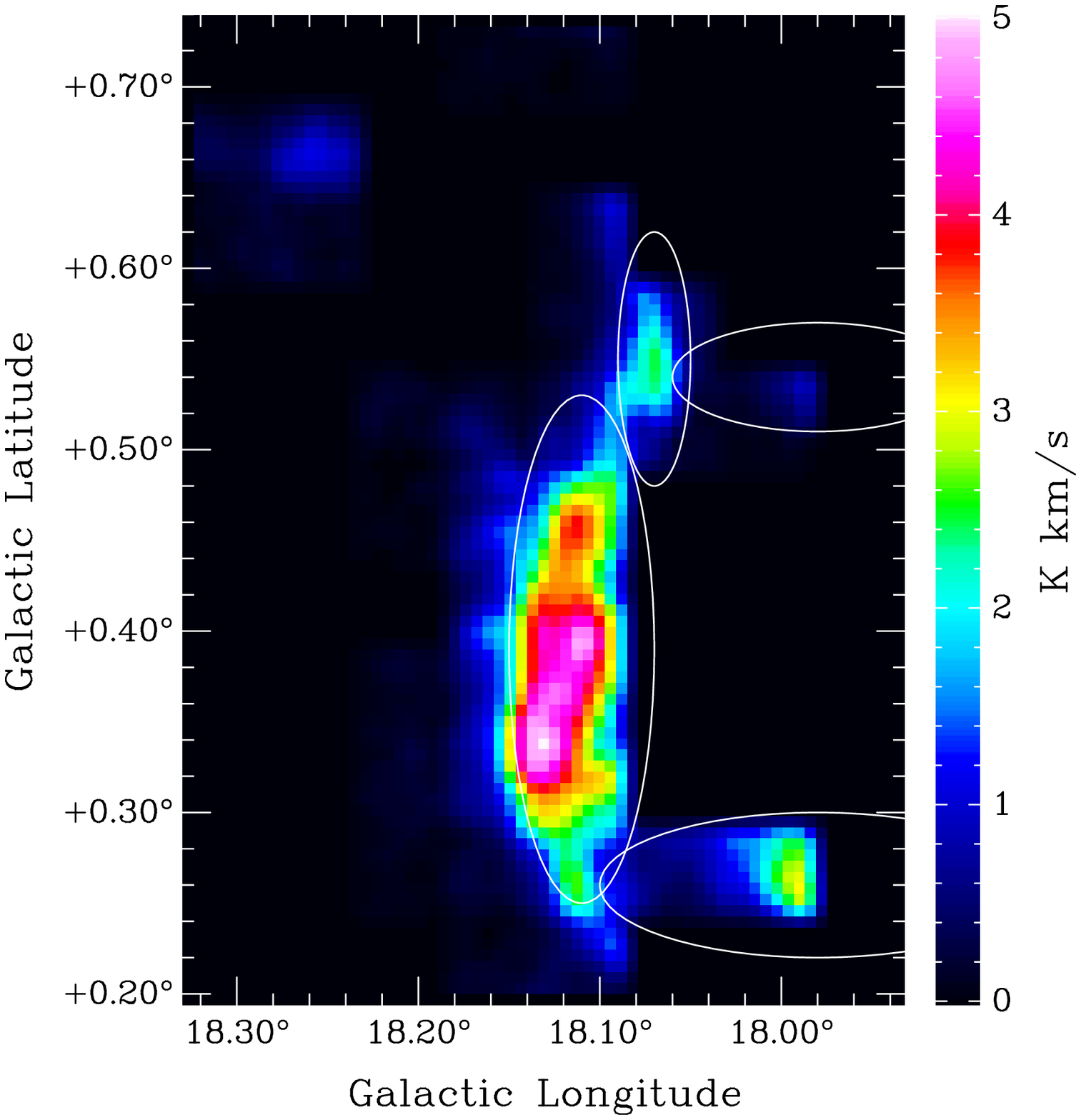}
\includegraphics[angle=-90,width=0.52\textwidth]{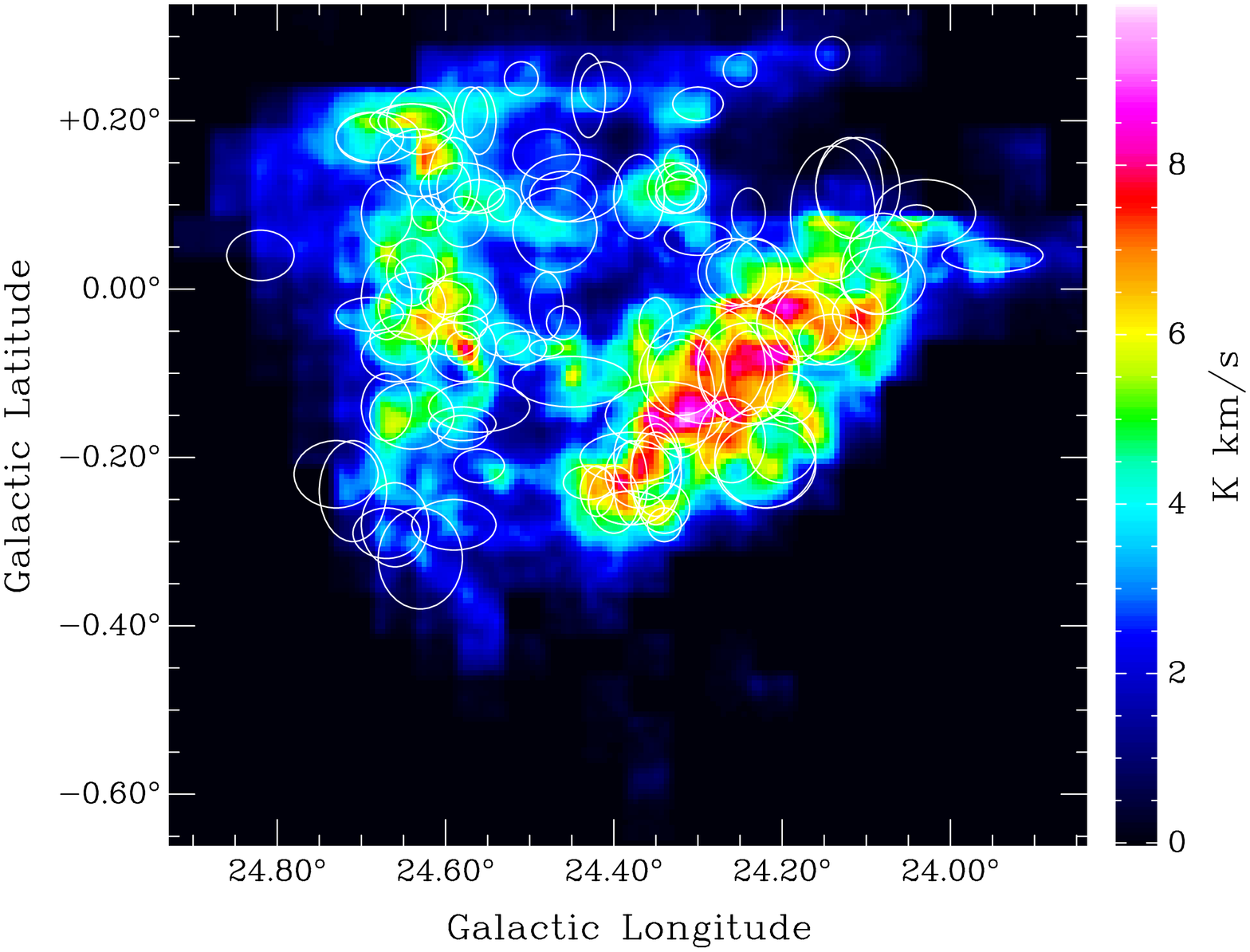}
\caption{\label{cloud1} GRSMC G018.14+00.39 (left) and GRSMC G043.34$-$00.36 (right)
  integrated intensity images.  Because it is very difficult to
  represent the asymmetric, three-dimensional output of CLUMPFIND in a
  two-dimensional image, we have include on these images white
  ellipses that indicate the approximate extent of each clump, with
  the major and minor axes equivalent to the projected extent of the
  clump in each direction. The identification of clumps is
  straightforward for GRSMC G018.14+00.39, but more complicated for
  GRSMC G043.34$-$00.36.}
\end{figure}

Figure~\ref{cloud1} shows the integrated intensity images for two
clouds identified in the GRS. The \lb\, images were produced by
integrating the emission over their velocity range, such that the
integrated intensity I = $\int$ \tmb\, $dv$. Because it is very
difficult to represent the asymmetric, three-dimensional output of
CLUMPFIND in a two-dimensional image, we have included on these images
white ellipses that indicate the approximate area of the region
identified by CLUMPFIND for each clump. To indicate the extent of the
clouds, we use two-dimensional ellipses whose major and minor axes are
equal to the projected extent of the clump in each direction. The
identification clumps is straightforward for GRSMC G018.14+00.39, but
more complicated for GRSMC G043.34$-$00.36 (see Fig.~\ref{cloud1}).

We find the number of clumps within these 829 molecular clouds range
from 1 to 111, with a typical molecular cloud containing $\sim$ 7
clumps. Because some clouds are too faint, we detect no clumps within
them above the clumpfind threshold. However, most of the clouds
($\sim$ 96\%) contain at least one clump that has a peak \tmb\, $>$ 10
times the average temperature of the cloud. These are the
brightest clumps in the clouds where star-formation will likely occur.


\section{Description of the catalog}

Table~\ref{cloud-list} gives a sample of the catalog entries for each
molecular cloud. Table~\ref{clump-list} gives a sample of the catalog
entries for the clumps. The complete catalogs are available in
electronic format at
http://www.bu.edu/galacticring/molecular\_clouds.html.

\subsection{Clouds}

For each molecular cloud identified, we list parameters output from
the CLUMPFIND algorithm in addition to several derived quantities. The
columns of Table~\ref{cloud-list} are as follows: (1) the molecular
cloud name, designated as GRSMC (for GRS Molecular Cloud) followed by
the Galactic longitude and latitude ($\ell$ and $b$) coordinates of
the peak of the emission in degrees, e.g.\ GRSMC G053.59+00.04; (2),
(3) and (4) the Galactic coordinates ($\ell$ and $b$) and velocity
(\vlsr) of the peak voxel of the emission; (5) the velocity FWHM
defined as 2.35 times the velocity dispersion ($\Delta$\,V); (6) main
beam temperature of the peak voxel (\tmb); (7) and (8) the Galactic
coordinates ($\ell$ and $b$) of the centroid of the integrated
intensity image; (9) and (10) the semimajor and semiminor axes (a
and b), determined from the elliptical Gaussian fit to the integrated
intensity image; (11) position angle of the fitted ellipse, measured
counter-clockwise from the positive longitude axis (PA); (12) the
projected two-dimensional area (A); (13) the average main beam
temperature of the cloud, defined as the average main beam temperature
of all the voxels (T$_{av}$); (14) the peak integrated intensity
(I$_{peak}$); (15) the total \tcont\, integrated intensity
(I$_{total}$\footnote{where I$_{total}$ = $\int \int$ \tmb\, d$v$
d$\Omega$}); (16) the peak \hh\, column density N(H$_{2}$); and (17)
a flag denoting if the cloud lies on one of the survey boundaries
('X', 'Y' or 'V' indicate that the cloud lies on the boundary in
Galactic longitude, Galactic latitude or velocity, respectively.)

Because molecular clouds have irregular shapes and complex
morphologies, the parameters listed in columns (7)--(11) are given
simply to estimate the clouds' approximate position, size, and
orientation.

\subsection{Clumps}

The clumps listed here represent the compact, dense substructures
within the clouds. Because we are interested in selecting the
brightest, densest regions within each of the clouds, not all clouds
have entries in this list; for the faintest clouds, we often detected
no clumps. To describe the shape of the clumps we simply use the
extent in Galactic longitude and latitude output directly from
CLUMPFIND (rather than the elliptical fitting that was performed for
the clouds).

The columns of Table~\ref{clump-list} are as follows: (1) the
molecular cloud name (from Table~\ref{cloud-list}); (2) the clump
number, e.g.\ c1, c2, c3, etc; (3), (4) and (5) the Galactic
coordinates ($\ell$ and $b$) and velocity (\vlsr) of the peak voxel of
the emission; (6), (7) and (8) the FWHM extent in longitude, latitude
and velocity ($\Delta \ell$, $\Delta b$, and $\Delta$\,V); (9) the
main beam temperature of the peak voxel (\tmb); (10) the projected
two-dimensional area (A); (11) the peak integrated intensity
(I$_{peak}$); (12) the total \tcont\, integrated intensity
(I$_{total}$); (13) the peak \hh\, column density (N(H$_{2}$)); and
(14) a flag denoting if the clump lies on a boundary.

\section{Ensemble properties}

\subsection{Galactic distribution}

\begin{figure}
\includegraphics[angle=-90,width=0.95\textwidth]{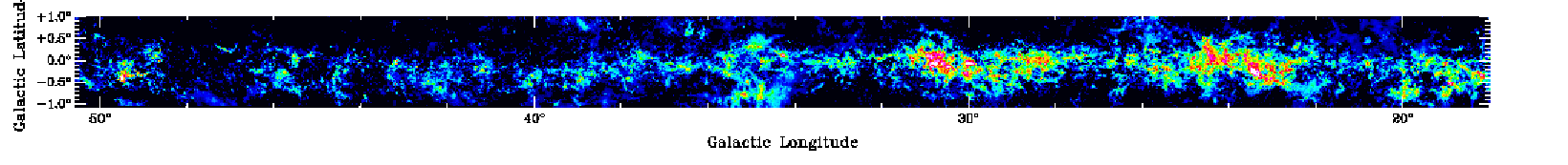}\\
\includegraphics[width=0.45\textwidth]{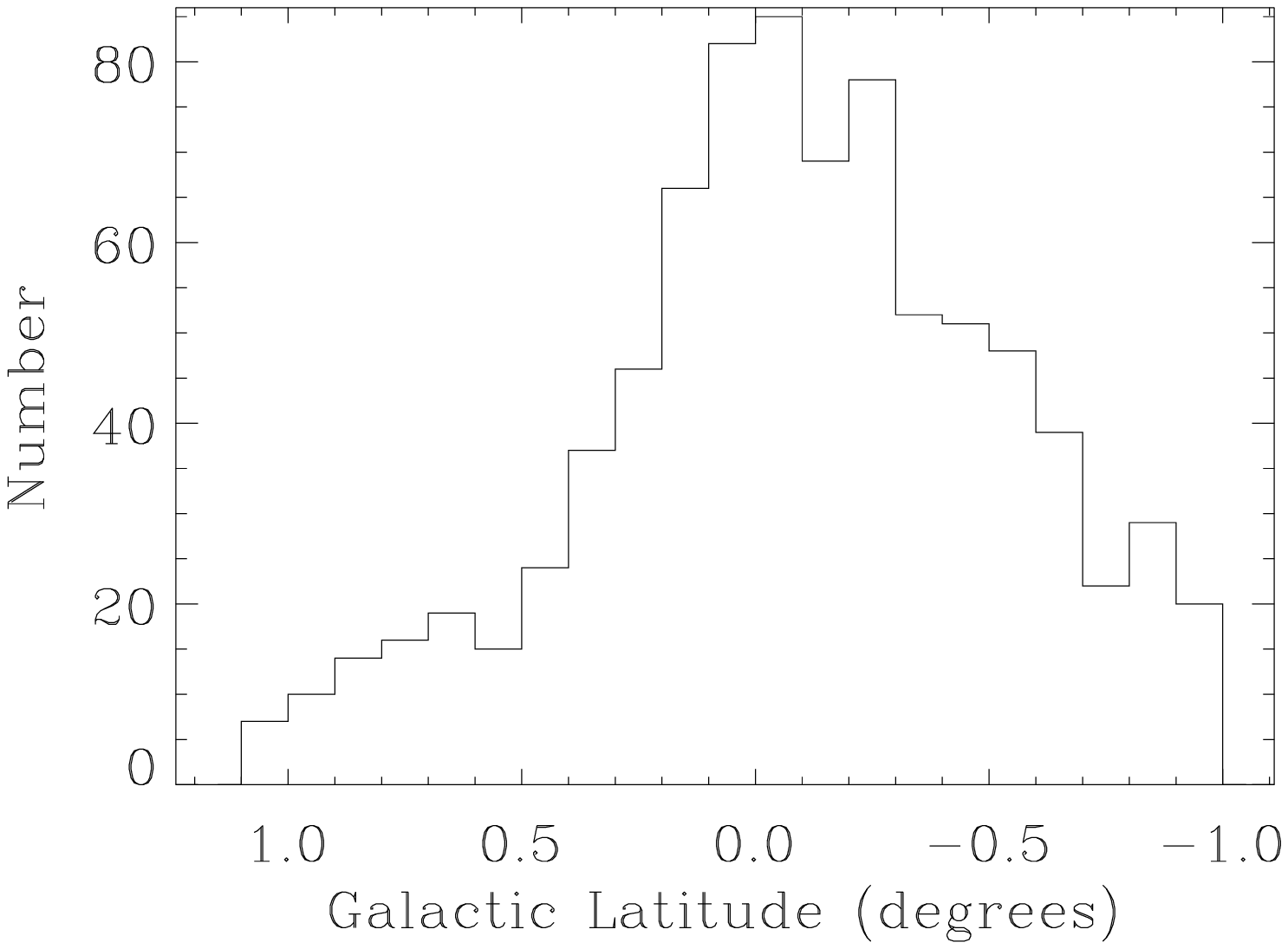}
\includegraphics[width=0.45\textwidth]{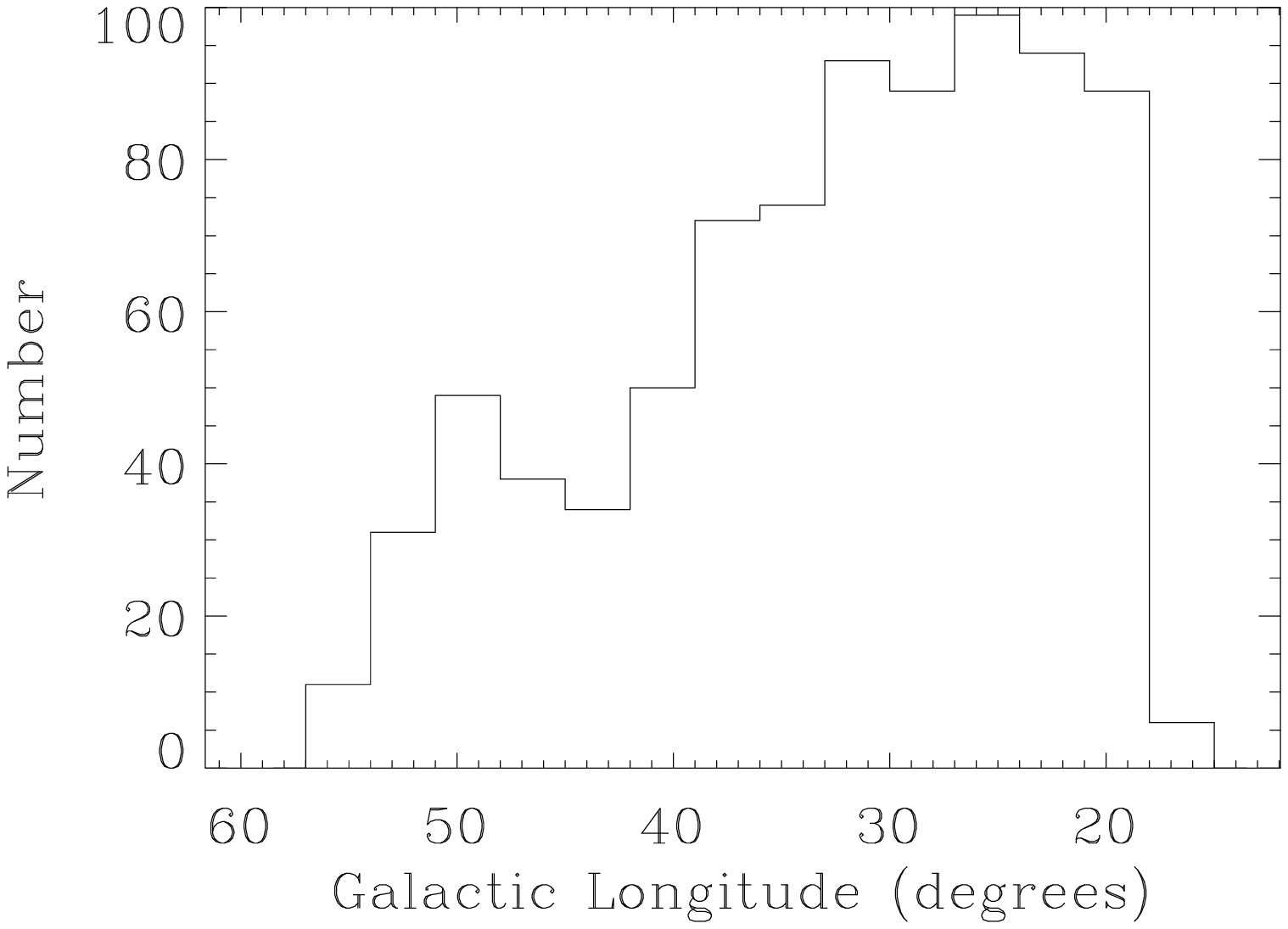}
\caption{\label{galdist} The Galactic distribution of the 829 molecular clouds 
    identified in the GRS. The top panel shows the integrated
    intensity image for the GRS (square root color scale, from 0 to 40
    \Kkms). The lower panels show the number distribution of molecular
    clouds in Galactic latitude (left) and longitude (right).}
\end{figure}

\begin{figure}
\includegraphics[angle=-90,width=0.9\textwidth]{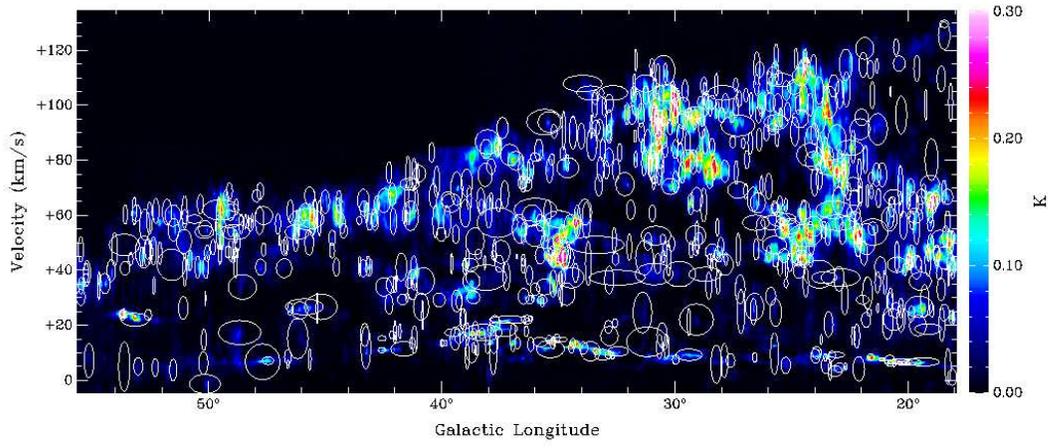}
\caption{\label{lv} The \lv\, diagram for the GRS. The color scale represents the 
            \tcont\, emission averaged over Galactic latitude.  The
            ellipses mark the approximate extent in Galactic longitude
            and velocity for each of the 829 clouds identified.}
\end{figure}

Figure~\ref{galdist} shows the number distribution in Galactic
latitude (left) and longitude (right) of the 829 molecular clouds
identified in the GRS.  Included in this figure for comparison is the
integrated intensity image for the GRS, that is the \tcont\, emission
integrated over all velocities. The location of the clouds in
\lv\, space is shown in Figure~\ref{lv}, overlaid on the GRS \lv\, image, 
which is the \tcont\, emission averaged over all Galactic latitudes. The
clouds identified span the complete range in Galactic longitude,
latitude, and velocity as covered by the GRS.

\begin{figure}
\includegraphics[width=0.5\textwidth]{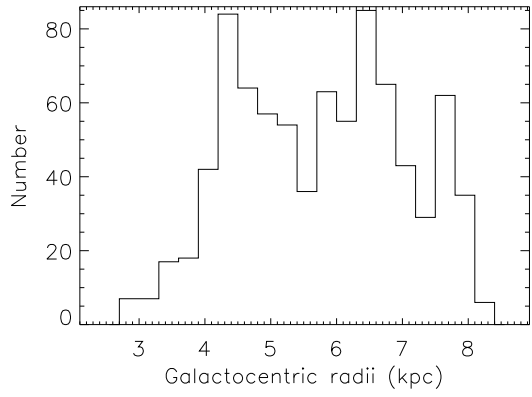}
\caption{\label{galrad} Galactocentric radial distribution of the molecular clouds 
   identified in the GRS.}
\end{figure}

Because the Milky Way rotation is well approximated by an axisymmetric
rotation curve, the line of sight velocity directly relates to a
Galactocentric radius. Thus, the \lv\, positions can uniquely
determine Galatocentric radii. To determine Galactocentric radii, we
assume circular motions and the rotation curve of \cite{Clemens85}
with (R$_{0}$,$\Theta_{0}$) = (8.5 kpc, 220\,\kms)\footnote{Note that
the Clemens rotation curve explicitly calls for a small velocity
correction factor due to a measured mis-calibration of the local
standard of rest.}.  Because we are sampling clouds primarily in the
flat part of the rotation curve, the derived kinematic distances are
insensitive to the exact choice of rotation curve.
Figure~\ref{galrad} shows the number distribution of clouds as a
function of Galactocentric radii. The peaks in this distribution
correspond to the known spiral features within the Galaxy.

In the inner Galaxy, a single Galactocentric radius corresponds to two
kinematic distances, one at the near distance and one at the far
distance. Accurate distances cannot be determined until this ambiguity
is resolved.  A comparison of GRS data with \hi\, self-absorption can
resolve this ambiguity and will be presented in a future paper
(J. Duval et al. in prep.)

\subsection{Properties of the clouds and clumps}

\begin{figure}
\includegraphics[width=0.5\textwidth]{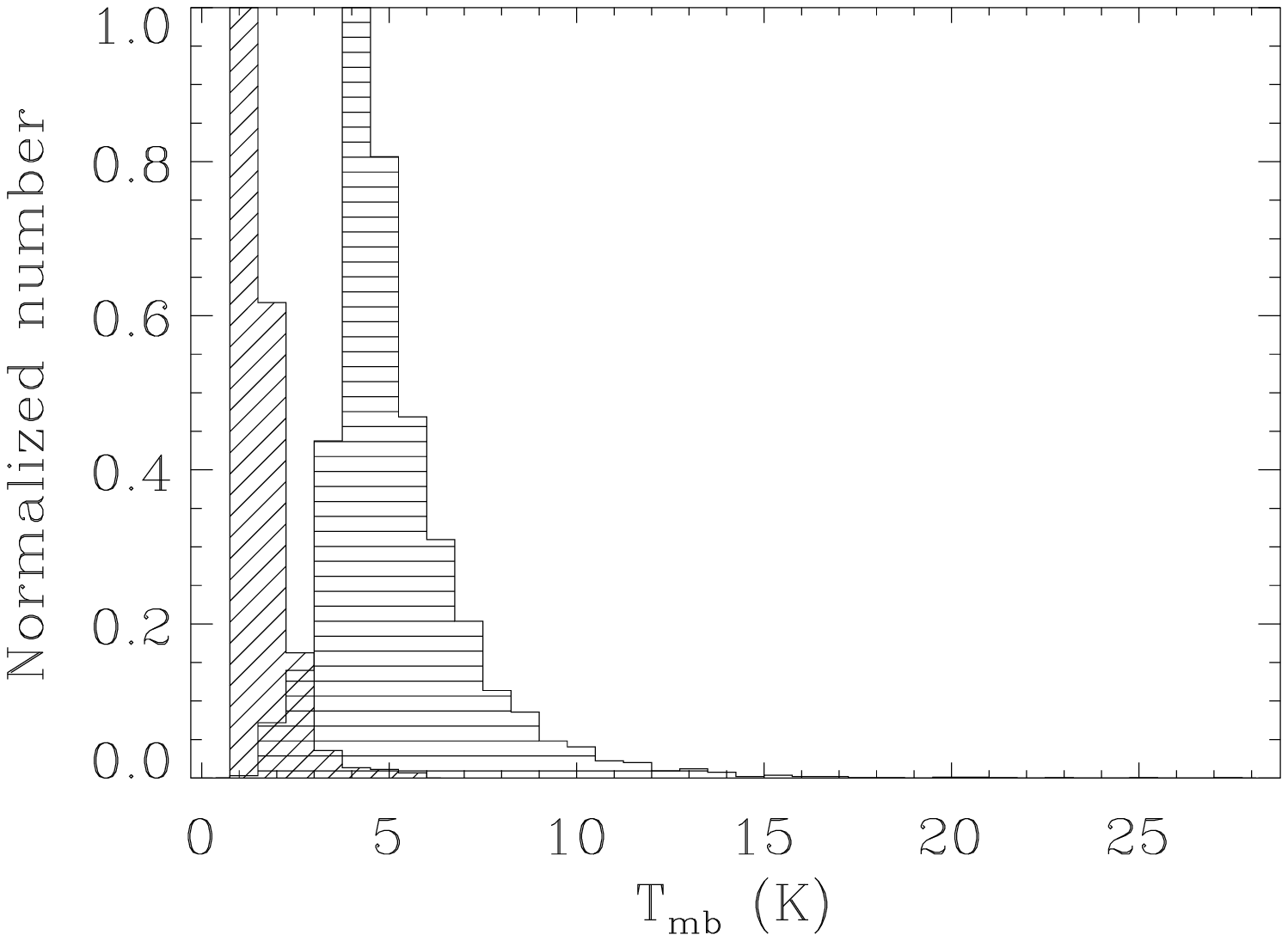}
\includegraphics[width=0.5\textwidth]{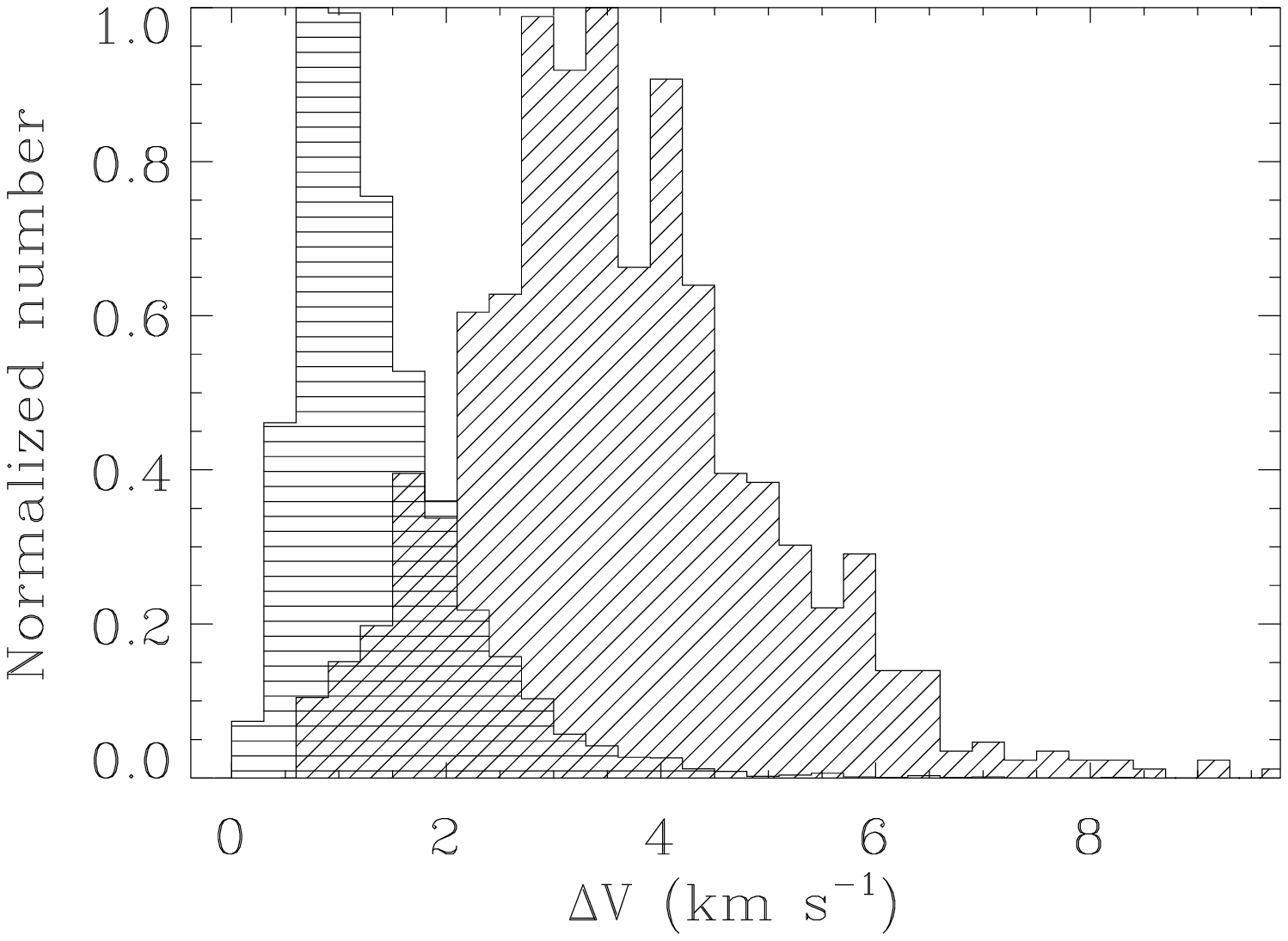}\\
\includegraphics[width=0.5\textwidth]{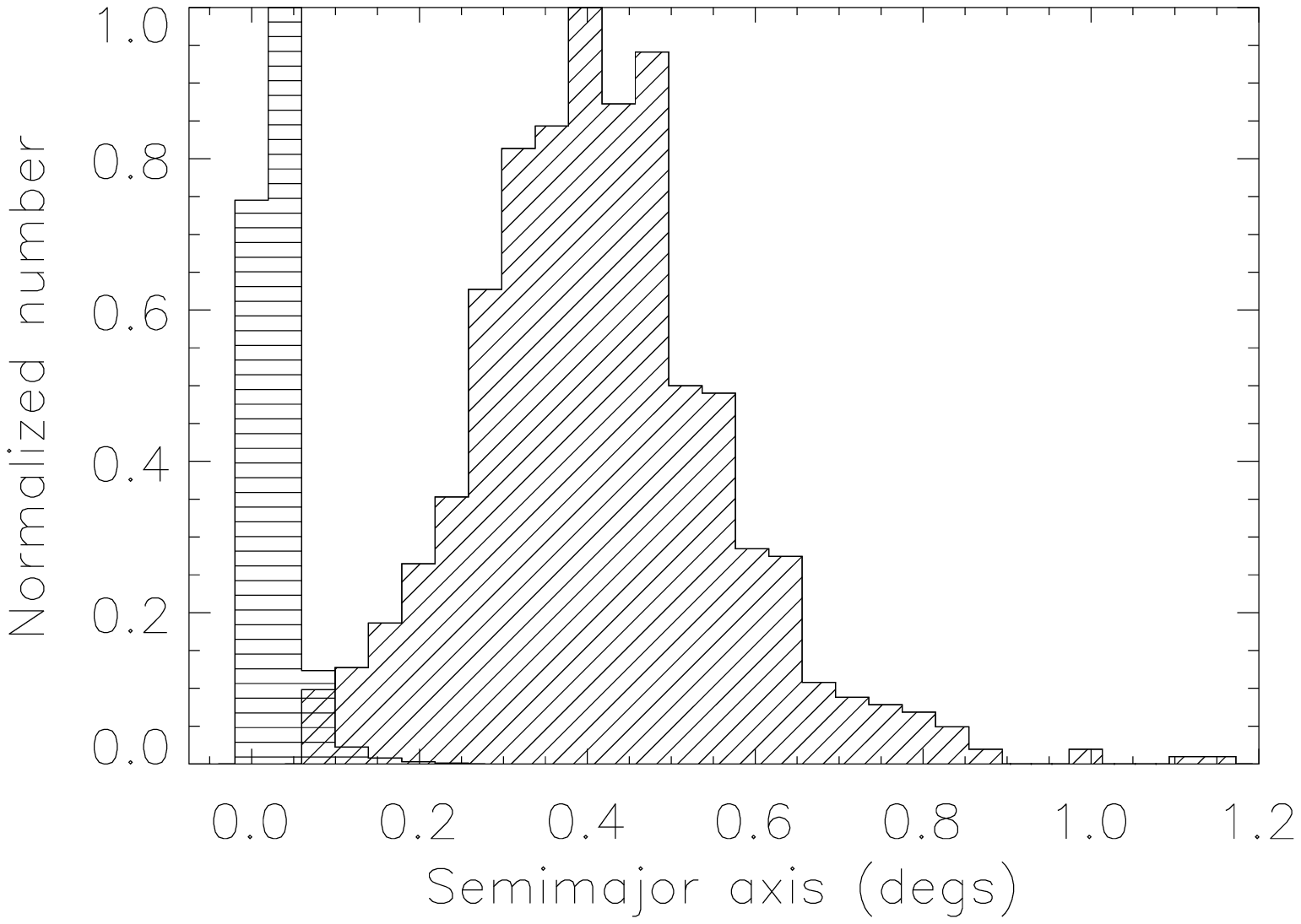}
\includegraphics[width=0.5\textwidth]{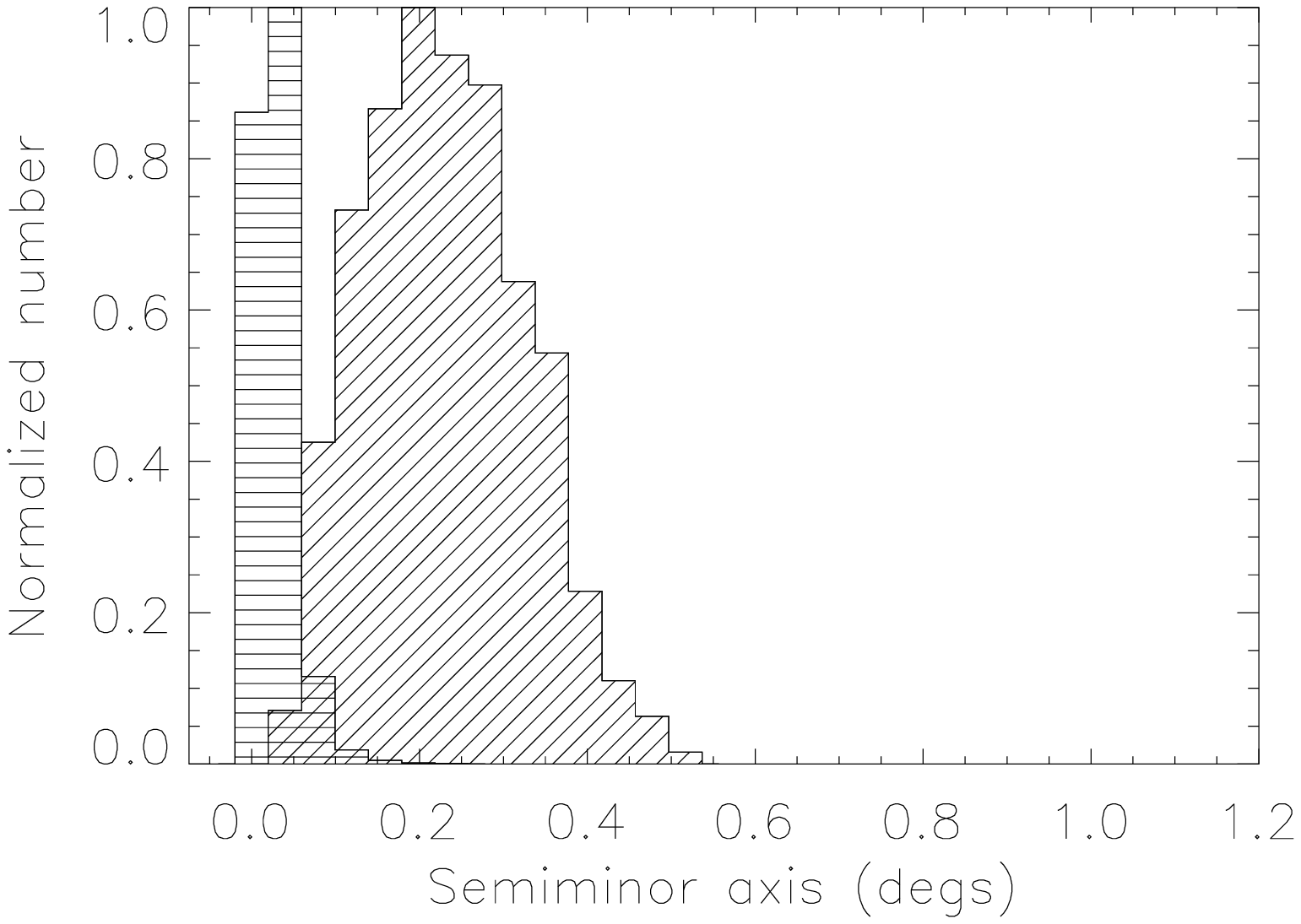}\\
\includegraphics[width=0.5\textwidth]{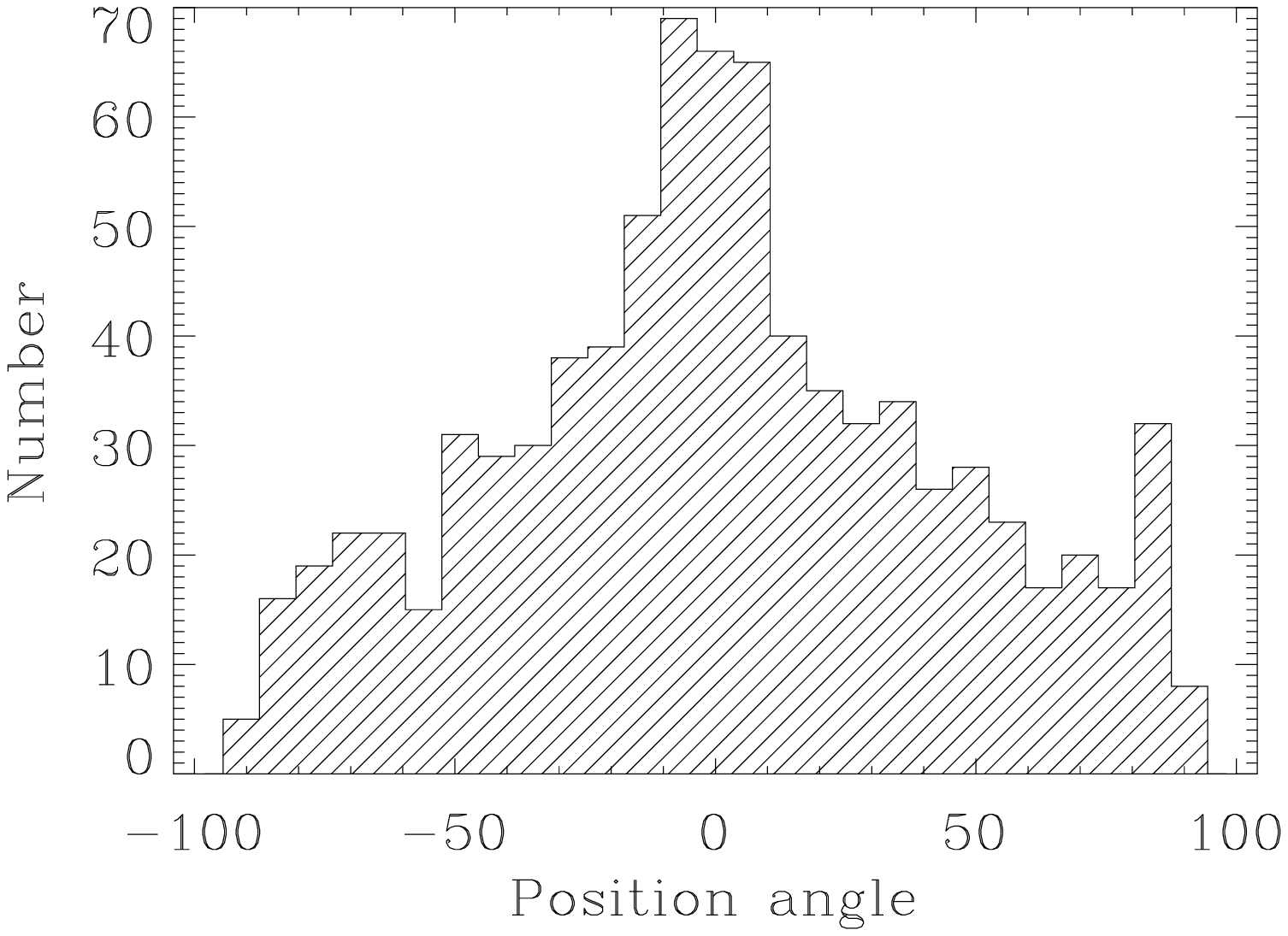}
\includegraphics[width=0.5\textwidth]{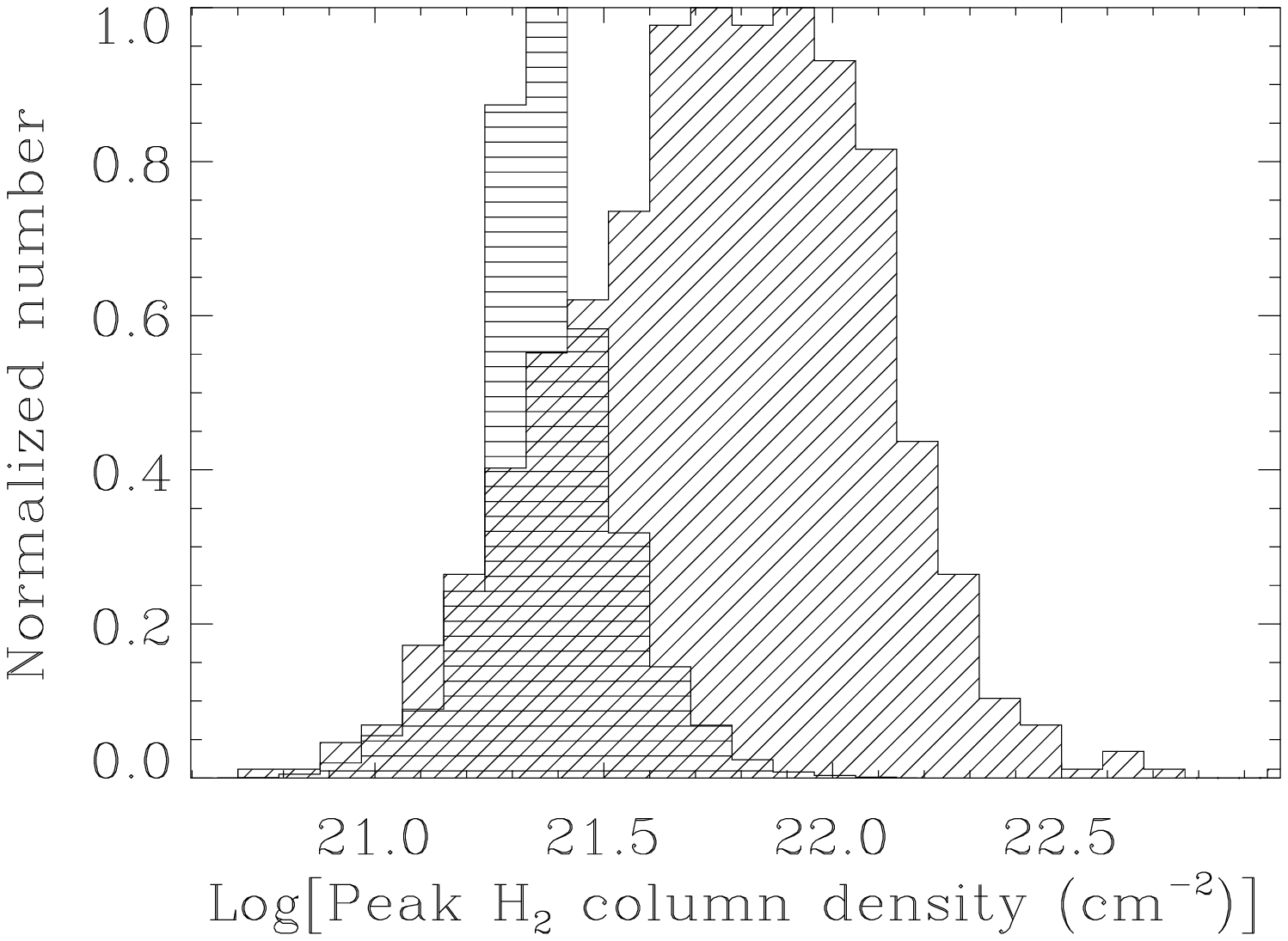}\\
\caption{\label{histograms}Normalized number distributions for the measured 
   parameters \tmb, line width, semimajor and semiminor axis, position angle,
   and peak N(\hh) for the clouds (diagonally hashed histogram) and
   clumps (horizontally hashed histogram). Position angles were only
   calculated for the clouds. See Table~\ref{grs-clouds} for a summary
   of these distributions.}
\end{figure}

Figure~\ref{histograms} shows the number distributions of the clouds
(diagonally hashed histograms) and clumps (horizontally hashed
histograms) as a function of peak \tmb, line width, semimajor axis,
semiminor axis, position angle, and peak \hh\, column density.

We find that the clouds have a mean peak \tcont\, \tmb\, of $\sim$
1.6\,K, $\Delta$V of $\sim$ 3.6\,\kms, semimajor axes of $\sim$ 0\fdg41,
and semiminor axes of $\sim$ 0\fdg23. The fact that the semimajor axes are
typically twice the semiminor axes suggests that most clouds are
elongated. In fact, the mean value of the ratio of semimajor to semiminor axis
is 2.0.  On the other hand, we find that the clumps are typically
brighter (peak \tmb\, of $\sim$ 5.2\,K), have lower line widths
($\Delta$V $\sim$ 1.4\,\kms), are smaller (semimajor axes of 0\fdg06), and
are rounder (the semimajor and semiminor axes are comparable).

The position angles measured for the clouds range from $-$90 to
90\degree\, with a peak value of 1\fdg2 (Fig.~\ref{histograms}). The
position angles were measured counter-clockwise from the positive
Galactic longitude axis, so a peak value in the distribution of 1\fdg2
implies most clouds are orientated along the Galactic plane. This
result is in agreement with previous studies using the GRS dataset
\citep{Koda06}. This result may arise from  an observational bias
due to the extended areal coverage of the GRS along the Galactic plane ($\sim$
38\degree\, were covered in Galactic longitude compared to the
2\degree\, covered in Galactic latitude). The fact that few clouds
were detected toward the edges of the coverage in Galactic latitude,
however, suggests that our results are not severely effected by this
lack of latitude coverage.

The peak \hh\, column density, N(\hh), was calculated by assuming
optically thin \tcont\, emission with an excitation temperature of 10K
and using standard conversion factors (see \citealp{Simon01}) via the
expression

\[N(H_{2}) = 4.92 \times 10^{20}\,\, I_{peak} \hspace{0.8cm} (cm^{-2}) \].

\noindent where $I_{peak}$ is the peak integrated intensity (\Kkms). 
We find that the clouds have a mean peak N(\hh) of $\sim$ 6.3 $\times$
10$^{21}$\,\cms\, while the clumps have slightly lower peak column
densities of $\sim$ 2.4 $\times$ 10$^{21}$\,\cms. This is opposite to
what is expected if the clumps are the densest parts of clouds where
star-formation will likely take place. However, this results is easily
explained by the fact that the clump distribution includes all clumps
within a cloud, makes the mean of the N(\hh) distribution lower for
the clumps compared to the clouds.

Table~\ref{grs-clouds} summarizes these parameters for the clouds and
clumps and lists the minimum, maximum, mean, median, standard
deviation, and slope of a power-law fit to the distributions.  We find
that the clouds identified have smaller average brightness
temperatures, larger line widths and sizes, and higher peak column
densities compared to the clumps.

\subsection{Determining the excitation temperatures and opacities}


To derive the excitation temperatures (\tex) and opacities ($\tau$) of
the clouds within our catalog, we make use of the \co\, University of
Massachusetts-Stony Brook survey (UMSB; \citealp{Sanders86}).  The
survey region covered the same region of the Galactic plane as the
GRS, but mapped the molecular clouds via their \cont\,
emission. Direct comparisons between the measured \cont\, and
\tcont\, temperatures are facilitated by the fact that the two surveys
were obtained with the same telescope (the FCRAO 14\,m). However,
because the UMSB survey was under-sampled (45\arcsec\, beam on a
3\arcmin\, grid; \citealp{Sanders86}) compared to the GRS (46\arcsec\,
beam on a 22\arcsec\, grid; \citealp{Jackson06}), the combination of
the two dataset can only provide rough average values for the entire
cloud.

Because the \co\, emission is typically optically thick, it can be used
to estimate the kinetic temperature of the gas via the expression

\[T_{\rm k} = 5.532 \left[ ln\left( 1+\frac{5.532}{T_{12}+0.837} \right) \right]^{-1} \hspace{0.8cm} (K)\]

\noindent where T$_{12}$ is the \cont\, brightness temperature 
measured from the UMSB survey.  The kinetic temperature can then be
used to calculate the opacities for the optically thin \tco\,
transition using

\[ \tau \approx \frac{k T_{mb}}{h \nu} \left[ \frac{1}{e^{\frac{h \nu}{k T_{k}}} -1} -  \frac{1}{e^{\frac{h \nu}{k T_{bg}}} -1} \right]^{-1}\]

\noindent where T$_{bg}$ is 2.7 K and \tmb\, is the \tcont\, brightness 
temperature measured within the cloud from the GRS. To calculate
T$_{12}$ and T$_{13}$ we calculate the mean \tmb\, from all the
corresponding voxels associated with each cloud as defined by
CLUMPFIND within the UMSB and GRS datasets respectively.  The
excitation temperature of the gas is then calculated using

\[T_{ex} = \frac{T_{mb}}{(1-e^{-\tau})} + T_{bg} \hspace{0.8cm} (K)\]

\begin{figure}
\includegraphics[width=0.5\textwidth]{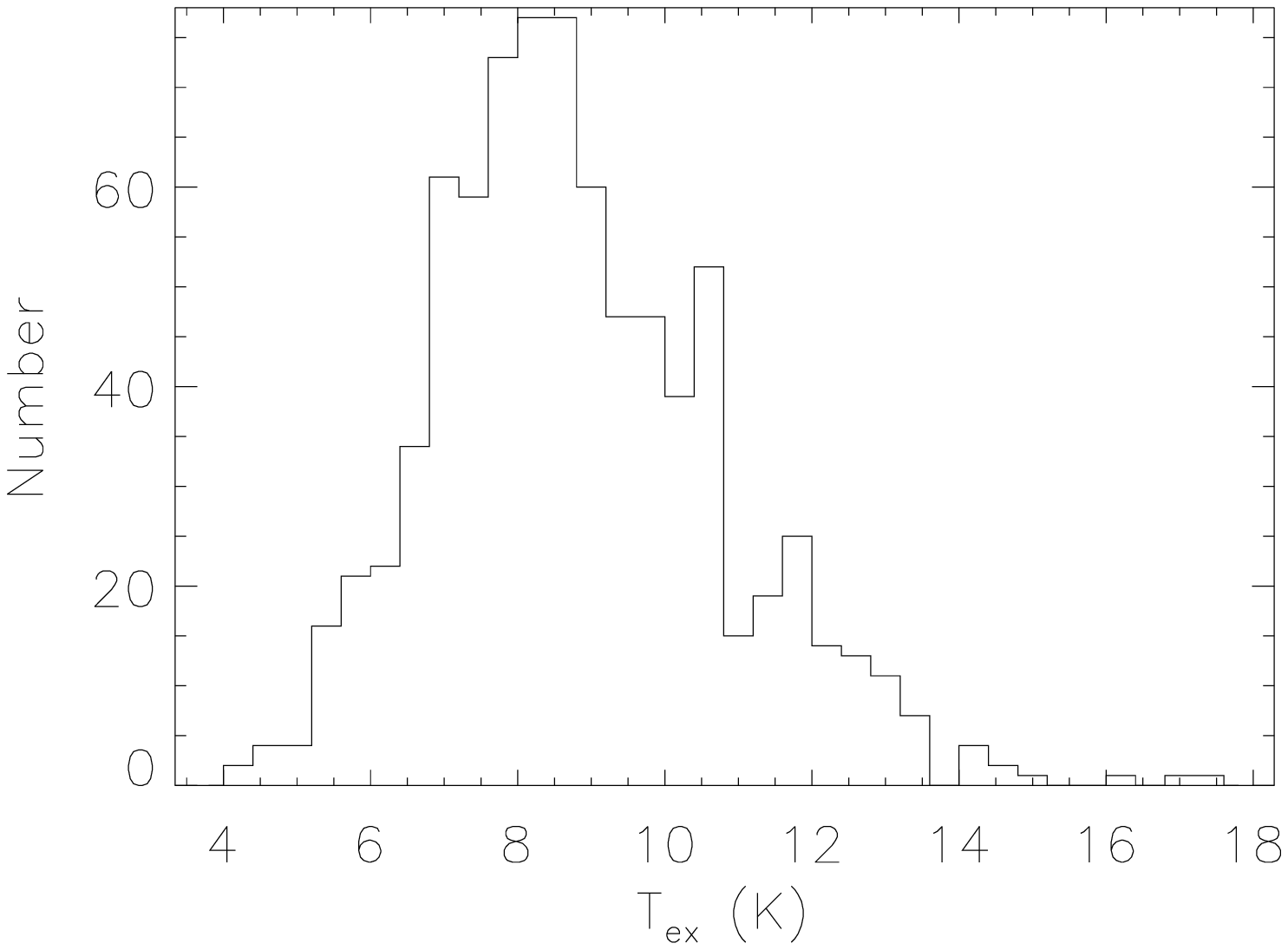}
\includegraphics[width=0.5\textwidth]{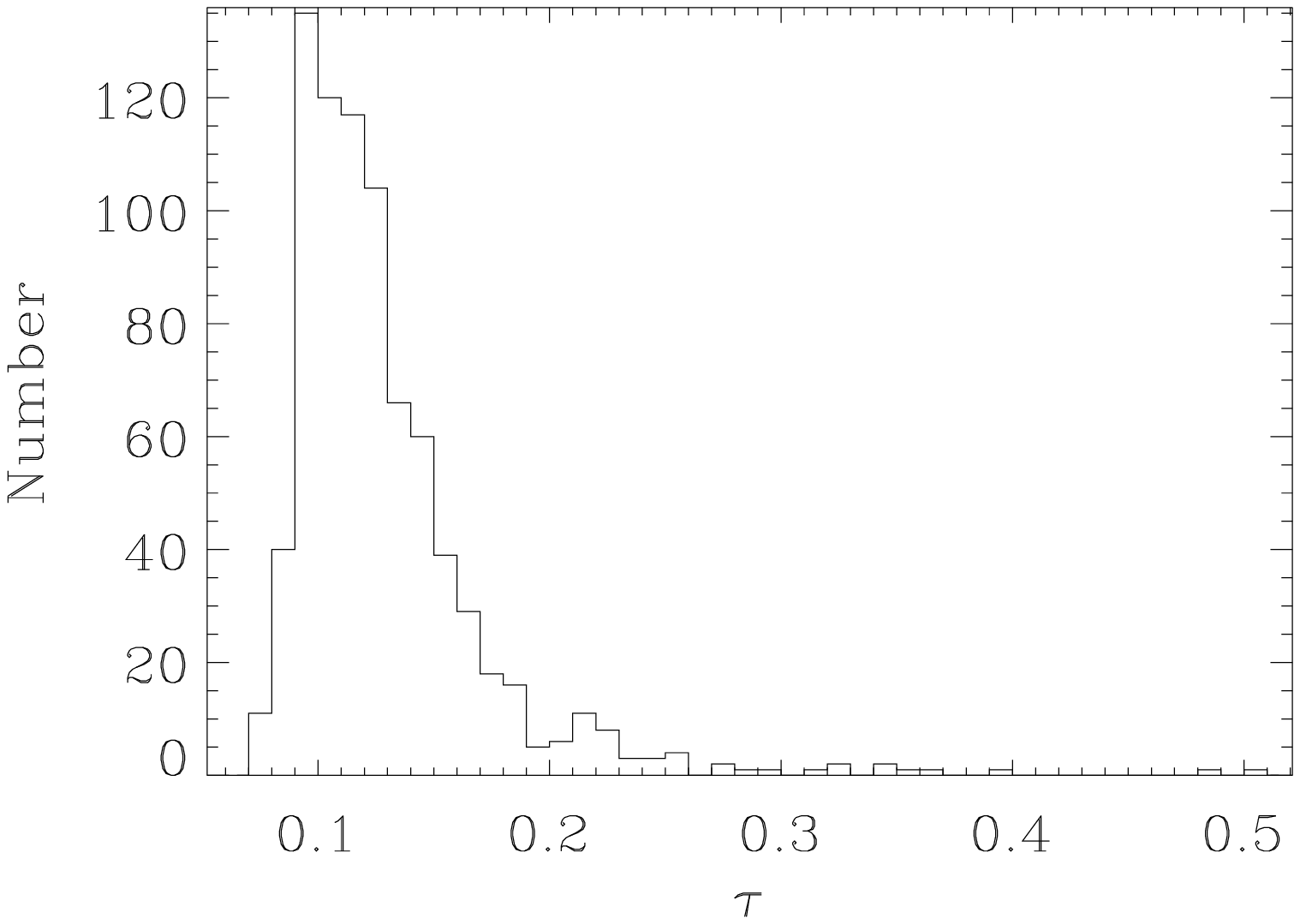}
\caption{\label{tex-tau} Number distributions 
   of the excitation temperature, \tex, (left) and the \tcont\,
   opacities, $\tau$, (right) for the clouds. These were calculated by
   combining the \tcont\, GRS emission with \co\, emission from the
   University of Massachusetts-Stony Brook survey (UMSB;
   \citealp{Sanders86}). The molecular clouds have mean \tex\, of
   $\sim$ 9 K and $\tau$ of 0.13.}
\end{figure}

Figure~\ref{tex-tau} shows the histograms of \tex\, and $\tau$ for the
clouds.  We find that the clouds typically have \tex\, of $\sim$ 9 K
and a \tcont\, $\tau$ of 0.13 (see Table~\ref{grs-clouds} for a
summary). Most clouds have low opacities, which confirms the fact that
the \tcont\, is optically thin and, thus, a good tracer of the column
density and mass.

\subsection{Properties of clouds inside and outside the 5 kpc ring}

With a large sample of molecular clouds we can now investigate their
properties in a range of Galactic environments. For instance, does the
temperature, line width, and column density of the molecular clouds
within active star-forming regions differ from those in more quiescent
regions?  Given that the majority of Galactic star formation occurs
within the 5~kpc molecular ring, one might expect to see differences
in the properties of the molecular clouds within the ring, compared to
clouds that lie outside the ring.

To test this idea we have separated the clouds identified with the GRS
into those that lie inside the 5 kpc molecular ring from those that
lie outside the ring. We have defined clouds that are associated with
the peak in the Galactocentric radius (R$_{G}$) distribution
(Fig.~\ref{galrad}) at $\sim$ 4.5 kpc as those associated with the
ring (i.e. all clouds with 4 kpc $<$ R$_{G}$ $<$ 5 kpc). Thus, the
number of clouds within the ring, N$_{in}$, is 206 (25\%), while the
number of clouds outside the ring, N$_{out}$, is 623 (75\%).

\begin{figure}
\includegraphics[width=0.4\textwidth]{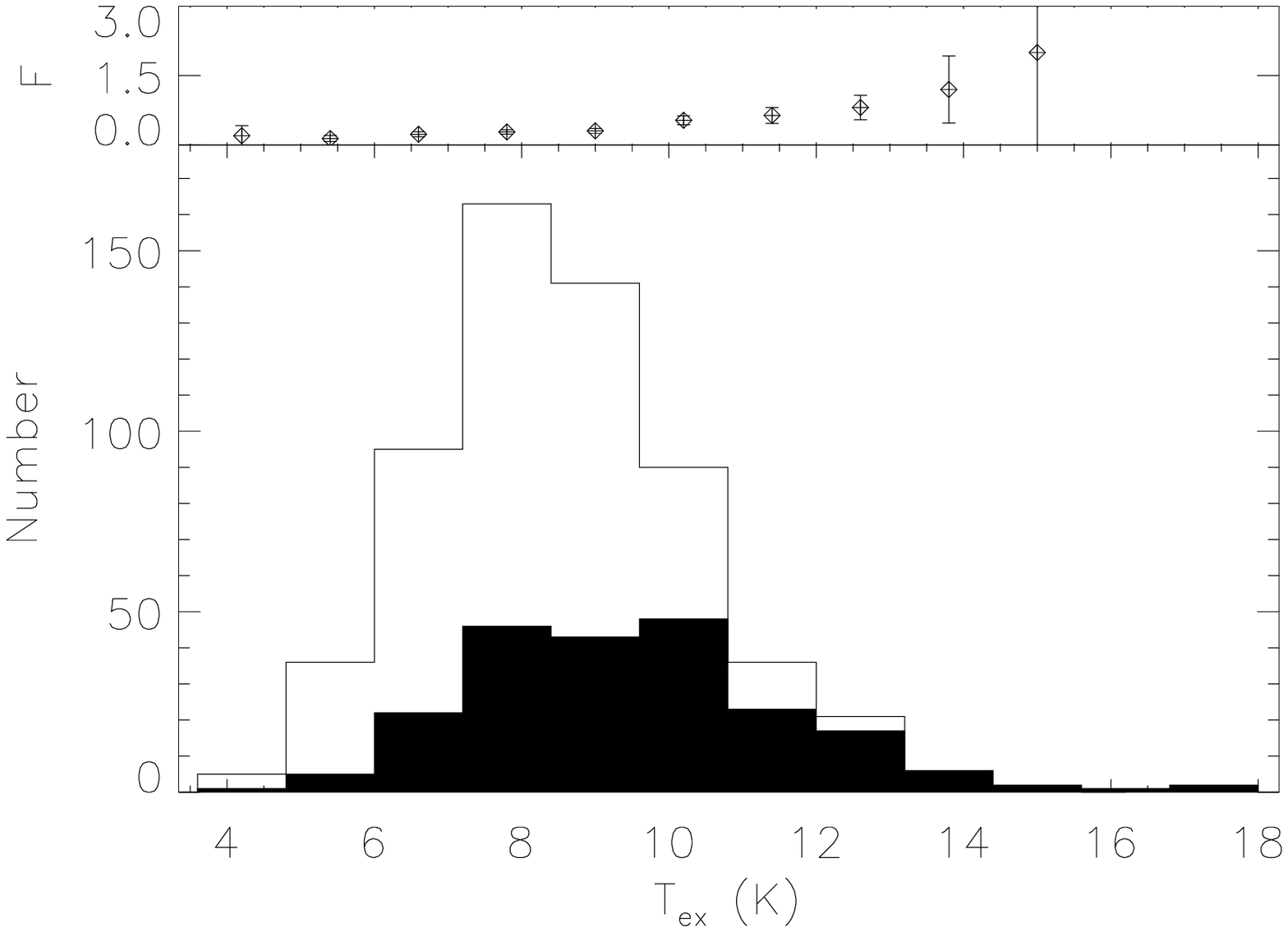}
\includegraphics[width=0.4\textwidth]{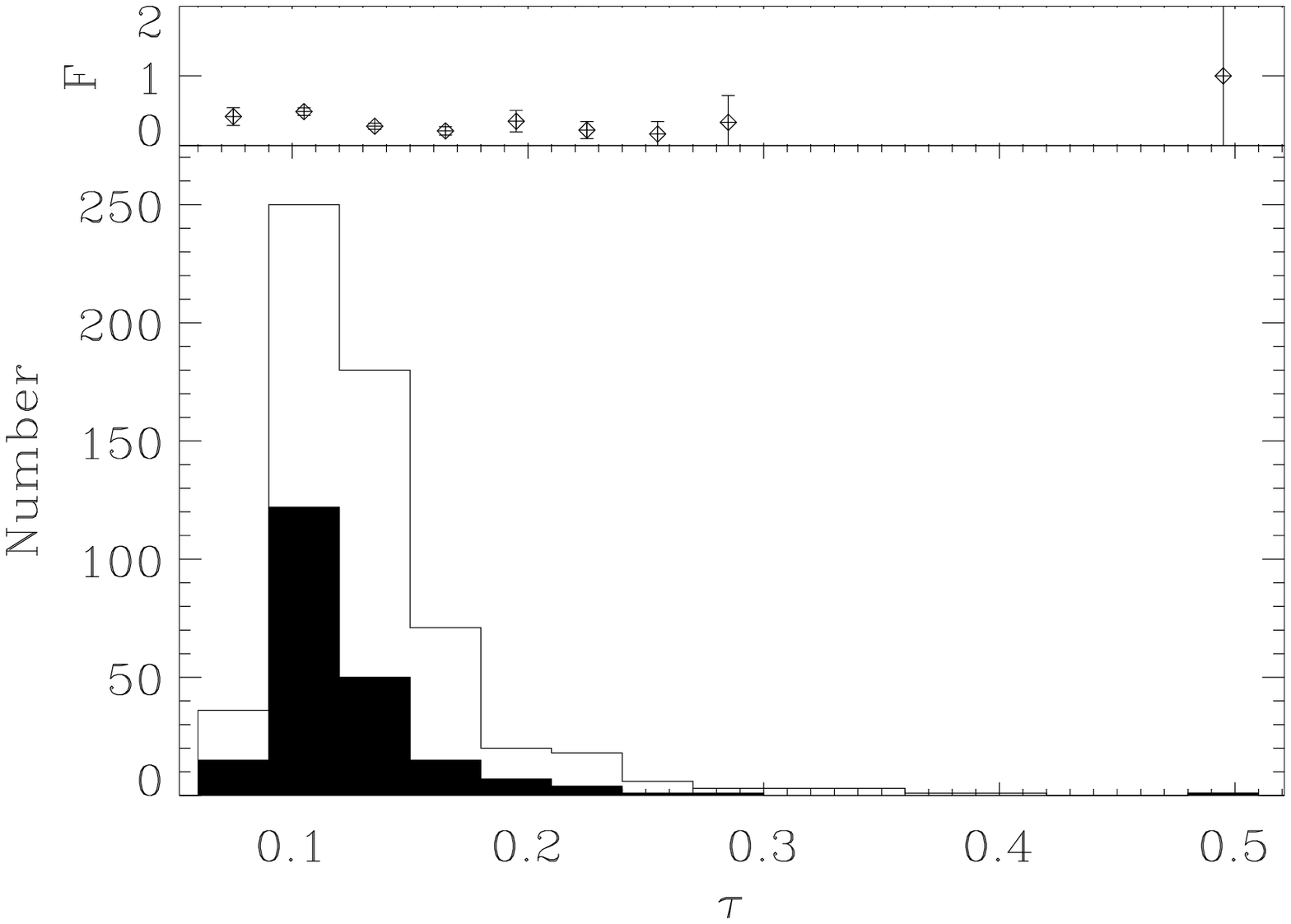}\\
\includegraphics[width=0.4\textwidth]{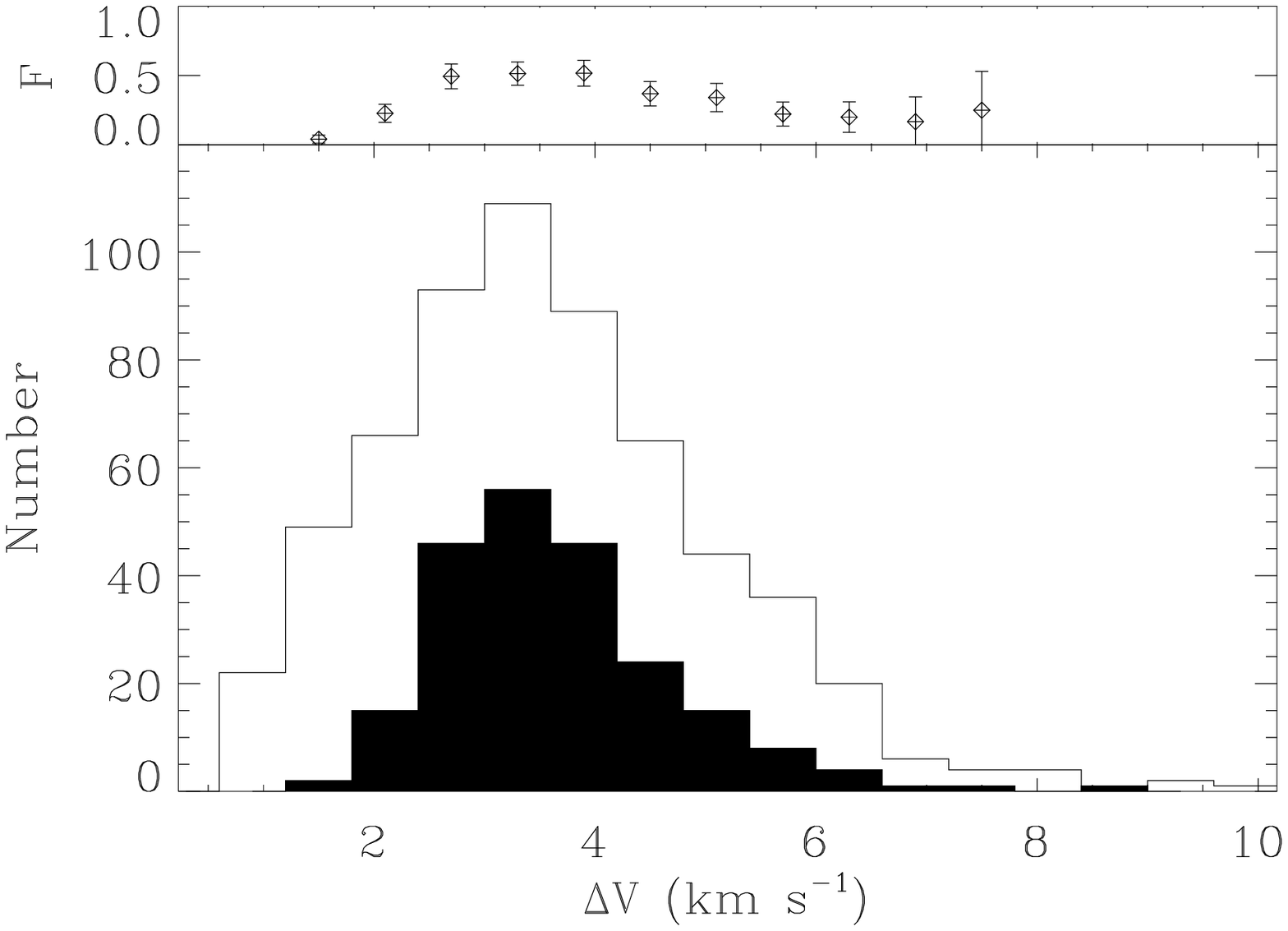}
\includegraphics[width=0.4\textwidth]{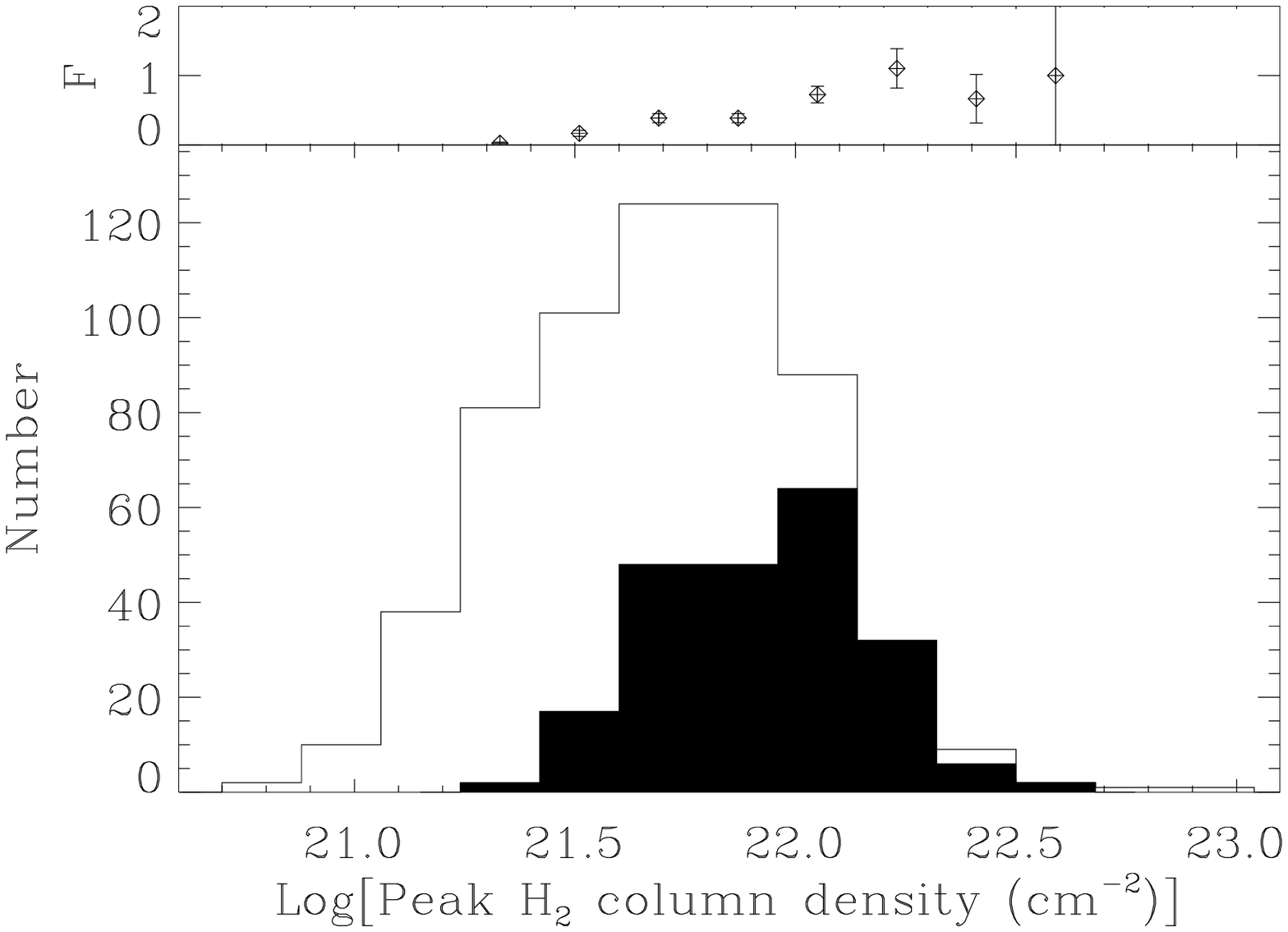}\\
\includegraphics[width=0.4\textwidth]{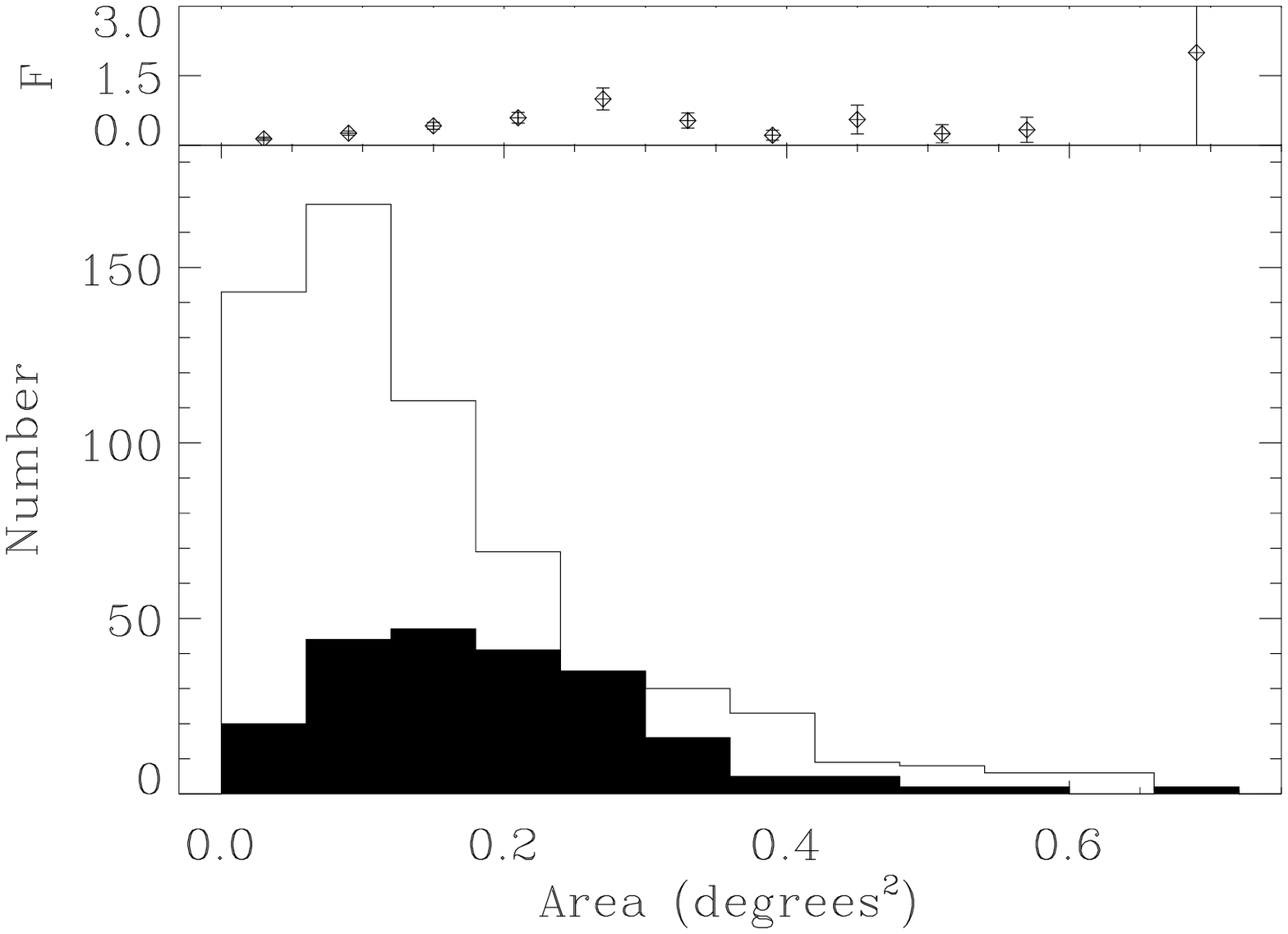}
\includegraphics[width=0.4\textwidth]{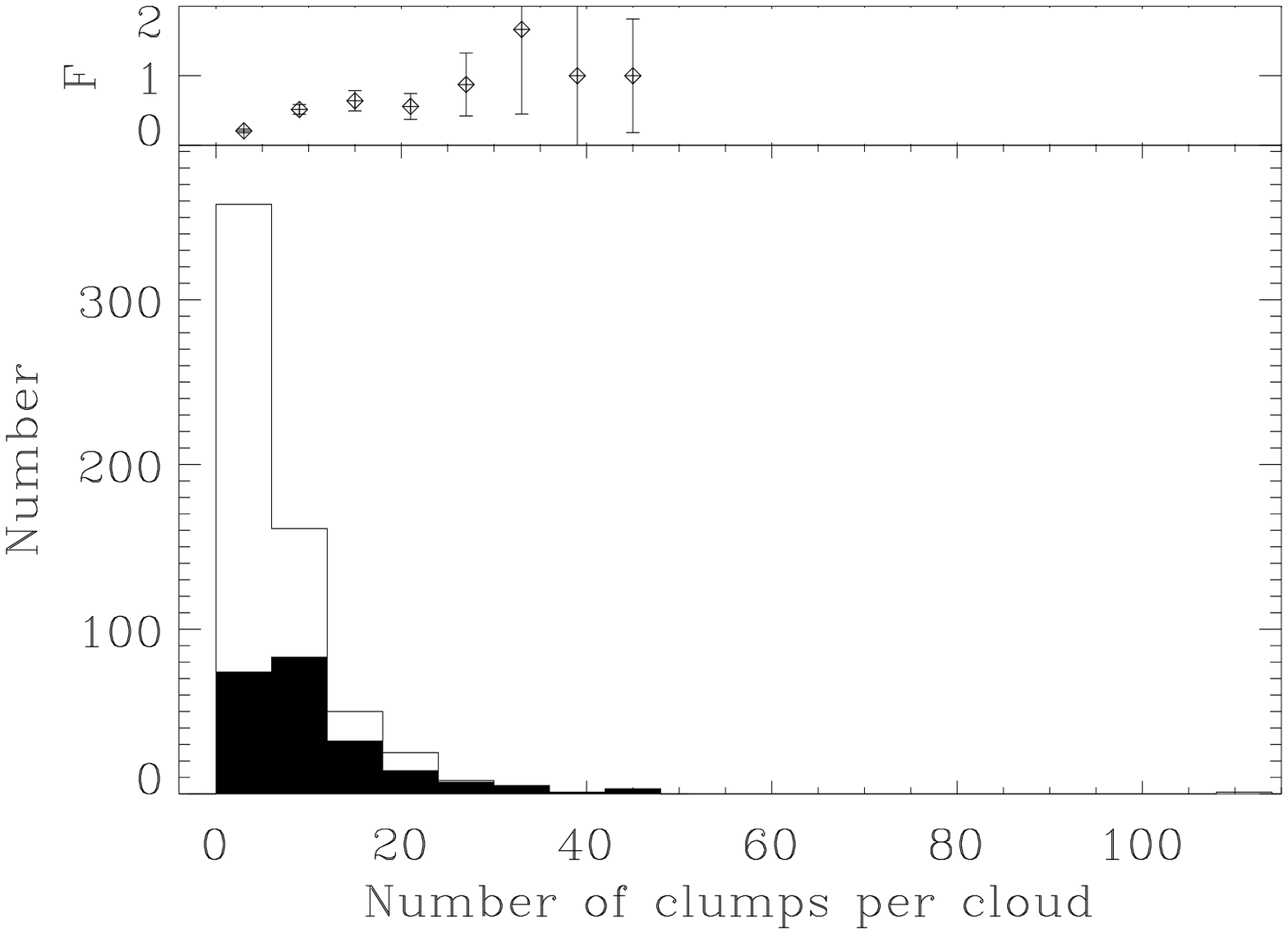}\\
\caption{\label{ring} Histograms of the parameters for clouds located in 
     the 5\,kpc molecular ring (solid histogram) and outside the ring
     (open histograms). The top panel in each plot shows the fraction
     of the number of clouds in the ring for each bin. We find that
     the clouds within the ring typically have warmer temperatures,
     higher column densities, larger areas, and more clumps compared
     to clouds located outside the ring. This is expected if these
     clouds are actively forming stars.}
\end{figure}

Figure~\ref{ring} shows the number of clouds in and out of the ring
(solid and open histograms respectively) plotted against their values
of \tex, $\tau$, line width, peak N(\hh), area, and number of clumps.
Included in the top panels of each of these plots is the fraction of
clouds within the ring in each of the bins. These plots show that
clouds within the ring typically have warmer temperatures, higher
column densities, larger areas, and more clumps compared to clouds
located outside the ring.  The mean opacities appear comparable
between clouds inside and outside the ring. We also find clouds within
the ring have similar mean values of the line width compared to those
clouds outside the ring.

In addition to the general cloud properties, we have also performed a
K-S test for each of the properties to determine if the distribution
in the samples inside and outside of the 5\,kpc ring are derived from
the same parent distribution. Table~\ref{grs-clouds} lists the results
of the K-S tests and shows that for all properties, with the exception
of position angles, the distributions are not derived from the same
parent distribution. Thus, there are clear, and significant
differences in the distributions of the derived properties for the
clouds inside and outside the 5\,kpc molecular ring.

Warmer temperatures and higher column densities are exactly what is
expected for active star-forming regions. All the clouds have
line widths much greater than the thermal line width for gas at their
derived \Tk\, (for \Tk = 10\,K, the thermal line width for CO is
0.13\,\kms) suggesting that they are dominated by turbulent motions.

The fact that we see more clumps within the clouds associated with the
5 kpc molecular ring also strengthens the idea that these clouds are
forming stars. This suggests that clouds within the ring are highly
fragmented and have many dense, warm regions where the star formation
can occur, in contrast to the more quiescent, smoother clouds outside the ring.


While these results suggest that clouds within the ring have different
properties to those clouds found outside the ring, we need to consider
that clouds within the ring suffer more from line blending and their
identification as separate features or merged clouds is more dependent
on the values input into CLUMPFIND. Moreover, the effects of distance
also need to be considered; we can more easily separate nearby clouds
as opposed to those at the far side of the Galaxy. Nevertheless, our
analysis suggests that the clouds within the ring are significantly
different from those outside the ring.

\section{Summary}

Using the CLUMPFIND algorithm we have identified a large sample of
clouds and clumps within the \tco\,BU--FCRAO Galactic Ring Survey. In
total, we identified 829 clouds and 6124 clumps. We find the cloud
properties are comparable to typical Giant Molecular Clouds, while the
clumps are similar to the regions where star and cluster formation
occur. Moreover, it appears that clouds lying within the 5 kpc ring
typically have warmer temperatures, higher column densities, larger
areas, higher densities, higher masses, and more clumps compared to
clouds located external to the ring. This difference supports the idea
that star formation occurs within the ring. We note, however, that
there are inherent difficulties in determining cloud properties
without the proper consideration of distance biases.

This catalog provides an invaluable tool for studies of molecular
clouds. For instance, establishing reliable kinematic distances to the
GRS molecular clouds and clumps, and to their embedded young stellar
objects and clusters, is essential in order to determine their masses,
sizes, distributions, and luminosities. Thus, combining the catalog of
GRS molecular clouds with IR Galactic plane surveys from {\it IRAS},
\MSX, 2MASS, and \Spitzer, one can obtain luminosities, masses, and
sizes of stars and clusters embedded within their natal molecular
material. One can then address the questions: What is is the spatial
distribution, luminosity function, and initial mass function of the
young stars forming both inside and outside the ring?  How does the
star-formation process within the ring differ from that at other
locations in the Galaxy, such as near the Sun, in nearby spiral arms,
and in the outer Galaxy?  Moreover, the internal structure of
molecular clouds, which traces the influence of turbulence in the
interstellar medium, can also be studied in a wide range of
star-forming environments.  With a large sample of clouds we can also
obtain their clump mass spectra and study its relation to the stellar
initial mass function. These are all important and outstanding
questions relating to Galactic star-formation.

\acknowledgments

This publication makes use of molecular line data from the Boston
University--FCRAO Galactic Ring Survey (GRS). The GRS is a joint
project of Boston University and Five College Radio Astronomy
Observatory, funded by the National Science Foundation under grants
AST-9800334, AST-0098562, AST-0100793, AST-0228993, \& AST-0507657.



\begin{thebibliography}{14}
\expandafter\ifx\csname natexlab\endcsname\relax\def\natexlab#1{#1}\fi

\bibitem[{{Burton}(1976)}]{Burton76}
{Burton}, W.~B. 1976, \araa, 14, 275

\bibitem[{{Cernicharo}(1991)}]{Cernicharo91}
{Cernicharo}, J. 1991, in NATO ASIC Proc. 342: The Physics of Star Formation
  and Early Stellar Evolution, 287

\bibitem[{{Clemens}(1985)}]{Clemens85}
{Clemens}, D.~P. 1985, ApJ, 295, 422

\bibitem[{{Clemens} {et~al.}(1988){Clemens}, {Sanders}, \&
  {Scoville}}]{Clemens88}
{Clemens}, D.~P., {Sanders}, D.~B., \& {Scoville}, N.~Z. 1988, ApJ, 327, 139

\bibitem[{Combes(1991)}]{Combes91}
Combes, F. 1991, ARA\&A, 29, 195

\bibitem[{{Dame} {et~al.}(2001){Dame}, {Hartmann}, \& {Thaddeus}}]{Dame01}
{Dame}, T.~M., {Hartmann}, D., \& {Thaddeus}, P. 2001, \apj, 547, 792

\bibitem[{{Goldsmith}(1987)}]{Goldsmith87}
{Goldsmith}, P.~F. 1987, in Astrophysics and Space Science Library, Vol. 134,
  Interstellar Processes, ed. D.~J. {Hollenbach} \& H.~A. {Thronson}, Jr.,
  51--70

\bibitem[{{Jackson} {et~al.}(2006){Jackson}, {Rathborne}, {Shah}, {Simon},
  {Bania}, {Clemens}, {Chambers}, {Johnson}, {Dormody}, {Lavoie}, \&
  {Heyer}}]{Jackson06}
{Jackson}, J.~M., {Rathborne}, J.~M., {Shah}, R.~Y., {Simon}, R., {Bania},
  T.~M., {Clemens}, D.~P., {Chambers}, E.~T., {Johnson}, A.~M., {Dormody}, M.,
  {Lavoie}, R., \& {Heyer}, M.~H. 2006, \apjs, 163, 145

\bibitem[{{Koda} {et~al.}(2006){Koda}, {Sawada}, {Hasegawa}, \&
  {Scoville}}]{Koda06}
{Koda}, J., {Sawada}, T., {Hasegawa}, T., \& {Scoville}, N.~Z. 2006, \apj, 638,
  191

\bibitem[{{Robinson} {et~al.}(1984){Robinson}, {Manchester}, {Whiteoak},
  {Sanders}, {Scoville}, {Clemens}, {McCutcheon}, \& {Solomon}}]{Robinson84}
{Robinson}, B.~J., {Manchester}, R.~N., {Whiteoak}, J.~B., {Sanders}, D.~B.,
  {Scoville}, N.~Z., {Clemens}, D.~P., {McCutcheon}, W.~H., \& {Solomon}, P.~M.
  1984, \apjl, 283, L31

\bibitem[{{Sanders} {et~al.}(1986){Sanders}, {Clemens}, {Scoville}, \&
  {Solomon}}]{Sanders86}
{Sanders}, D.~B., {Clemens}, D.~P., {Scoville}, N.~Z., \& {Solomon}, P.~M.
  1986, \apjs, 60, 1

\bibitem[{{Scoville} \& {Solomon}(1975)}]{Scoville75}
{Scoville}, N.~Z. \& {Solomon}, P.~M. 1975, \apjl, 199, L105

\bibitem[{{Simon} {et~al.}(2001){Simon}, {Jackson}, {Clemens}, {Bania}, \&
  {Heyer}}]{Simon01}
{Simon}, R., {Jackson}, J.~M., {Clemens}, D.~P., {Bania}, T.~M., \& {Heyer},
  M.~H. 2001, ApJ, 551, 747

\bibitem[{{Williams} {et~al.}(1994){Williams}, {de Geus}, \&
  {Blitz}}]{Williams94}
{Williams}, J.~P., {de Geus}, E.~J., \& {Blitz}, L. 1994, \apj, 428, 693

\end{thebibliography}

\clearpage
\clearpage
\clearpage
\clearpage
\clearpage
\clearpage
\clearpage


\clearpage
\begin{table}
\caption{\label{properties}Properties of Giant Molecular clouds, clumps, and cores \citep{Goldsmith87,Cernicharo91}.}
\centering
\begin{tabular}{lc|ccc}
\tableline
\tableline
\multicolumn{1}{c}{Properties} 
                  & & GMC                & Clump              &  Core \\
\tableline           
Size (pc)         & & 20--60             &  3--20             & 0.5--3 \\
Density (\cmc)    & & 100--300           & 10$^3$--10$^{4}$   & 10$^{4}$--10$^{6}$ \\
Mass (\Msun)      & & 10$^{4}$--10$^{6}$ & 10$^{3}$--10$^{4}$ & 10--10$^{3}$ \\
Linewidth (\kms)  & & 6--15              &  4--12             & 1--3\\
Temperature (K)   & & 7--15              &  15--40            & 30--100\\
\tableline
\end{tabular}
\end{table}
\clearpage
\begin{sidewaystable}
\caption{\label{cloud-list} Properties of the molecular clouds identified in the GRS.}
\scriptsize
\begin{tabular}{ccrrcccrccrcccccc}\hline\hline
Cloud & \multicolumn{2}{c}{Peak} & \multicolumn{1}{c}{V$_{LSR}$} & $\Delta$V & T$_{mb}$ & \multicolumn{2}{c}{Centroid} & a & b & \multicolumn{1}{c}{PA} & A & T$_{av}$ & I$_{peak}$ & I$_{total}$  & N(H$_{2}$) & Flag\\
GRSMC & $\ell$ & \multicolumn{1}{c}{$b$} & & & & $\ell$ & \multicolumn{1}{c}{$b$} & & & & & & & & & \\
\tiny  
& (\degree) & (\degree) & \multicolumn{1}{c}{(km s$^{-1}$)} & (km s$^{-1}$) & (K) & (\degree) & (\degree) & (\degree) & (\degree) & (\degree) & (deg$^{2}$) & (K) & (\Kkms) & (\Kkms\,deg$^{2}$) & ($\times$10$^{22}$ cm$^{-2}$) &  \\
\scriptsize
(1) & (2) & \multicolumn{1}{c}{(3)} & \multicolumn{1}{c}{(4)} & (5) & (6) & (7) & \multicolumn{1}{c}{(8)} & (9) & (10) & \multicolumn{1}{c}{(11)} & (12) & (13) & (14) & (15) & (16) & (17)\\
\hline
G053.59+00.04 & 53.59 & 0.04 &  23.7 &  1.99 & 5.75 & 53.69 & 0.01 &  0.52 &  0.19 &  16 & 0.27 & 1.90 &  36.5 & 0.95 & 1.8 & -  \\
G029.89$-$00.06 & 29.89 & $-$0.06 & 100.7 &  5.09 & 5.46 & 29.99 & $-$0.17 &  0.44 &  0.35 & -46 & 0.46 & 2.49 &  87.6 & 3.12 & 4.3 & -  \\
G049.49$-$00.41 & 49.49 & $-$0.41 &  56.9 &  9.77 & 5.25 & 49.57 & $-$0.39 &  0.38 &  0.23 & -21 & 0.18 & 2.18 & 193.2 & 1.64 & 9.5 & Y  \\
G018.89$-$00.51 & 18.89 & $-$0.51 &  65.8 &  2.80 & 5.17 & 18.80 & $-$0.56 &  0.31 &  0.31 & -47 & 0.29 & 2.31 &  57.5 & 1.51 & 2.8 & -  \\
G030.49$-$00.36 & 30.49 & $-$0.36 &  12.3 &  4.56 & 4.98 & 30.66 & $-$0.39 &  0.47 &  0.24 & -28 & 0.22 & 1.70 &  17.5 & 0.50 & 0.9 & -  \\
G035.14$-$00.76 & 35.14 & $-$0.76 &  35.2 &  5.00 & 4.92 & 35.22 & $-$0.78 &  0.31 &  0.22 & -11 & 0.20 & 2.36 &  59.3 & 1.90 & 2.9 & Y  \\
G034.24+00.14 & 34.24 & 0.14 &  57.8 &  5.98 & 4.81 & 34.19 & 0.05 &  0.50 &  0.41 &  -3 & 0.52 & 1.46 &  80.5 & 3.15 & 4.0 & -  \\
G019.94$-$00.81 & 19.94 & $-$0.81 &  42.9 &  2.81 & 4.58 & 19.97 & $-$0.80 &  0.42 &  0.18 &  -7 & 0.22 & 1.76 &  32.1 & 1.00 & 1.6 & Y  \\
G023.44$-$00.21 & 23.44 & $-$0.21 & 101.1 &  5.75 & 4.40 & 23.36 & 0.02 &  0.64 &  0.48 &  29 & 0.63 & 1.65 &  63.2 & 3.71 & 3.1 & -  \\
G038.94$-$00.46 & 38.94 & $-$0.46 &  41.6 &  2.97 & 4.33 & 39.01 & $-$0.51 &  0.36 &  0.28 & -10 & 0.23 & 1.90 &  31.8 & 1.02 & 1.6 & -  \\
G023.44$-$00.21 & 23.44 & $-$0.21 & 103.7 &  3.44 & 4.23 & 23.55 & $-$0.27 &  0.35 &  0.23 &  28 & 0.12 & 2.13 &  51.7 & 0.49 & 2.5 & -  \\
G030.79$-$00.06 & 30.79 & $-$0.06 &  94.7 &  6.12 & 4.23 & 30.86 & $-$0.04 &  0.39 &  0.28 &  66 & 0.32 & 2.24 &  79.5 & 2.68 & 3.9 & -  \\
G030.29$-$00.21 & 30.29 & $-$0.21 & 104.5 &  3.01 & 3.92 & 30.36 & $-$0.18 &  0.39 &  0.23 &  66 & 0.26 & 1.66 &  40.0 & 0.98 & 2.0 & -  \\
G053.14+00.04 & 53.14 & 0.04 &  22.0 &  2.39 & 3.88 & 53.15 & 0.09 &  0.42 &  0.26 &  74 & 0.29 & 2.01 &  26.8 & 0.93 & 1.3 & -  \\
G022.44+00.34 & 22.44 & 0.34 &  84.5 &  2.81 & 3.62 & 22.51 & 0.30 &  0.41 &  0.23 &  32 & 0.14 & 1.52 &  29.0 & 0.42 & 1.4 & -  \\
G024.49+00.49 & 24.49 & 0.49 & 102.4 &  5.24 & 3.58 & 24.48 & 0.30 &  0.52 &  0.32 &   8 & 0.46 & 1.46 &  67.8 & 2.21 & 3.3 & -  \\
G049.39$-$00.26 & 49.39 & $-$0.26 &  50.9 &  3.54 & 3.54 & 49.44 & $-$0.23 &  0.30 &  0.18 &  63 & 0.15 & 1.93 &  54.1 & 0.90 & 2.7 & -  \\
G019.39$-$00.01 & 19.39 & $-$0.01 &  26.7 &  3.88 & 3.48 & 19.53 & 0.02 &  0.42 &  0.26 &  -2 & 0.31 & 1.73 &  39.0 & 2.00 & 1.9 & -  \\
G034.74$-$00.66 & 34.74 & $-$0.66 &  46.7 &  4.33 & 3.46 & 34.88 & $-$0.63 &  0.42 &  0.34 &  51 & 0.34 & 1.91 &  35.3 & 2.27 & 1.7 & Y  \\
G023.04$-$00.41 & 23.04 & $-$0.41 &  74.3 &  4.20 & 3.40 & 23.07 & $-$0.45 &  0.50 &  0.31 &   7 & 0.45 & 1.54 &  45.9 & 2.31 & 2.3 & -  \\
G018.69$-$00.06 & 18.69 & $-$0.06 &  45.4 &  3.86 & 3.33 & 18.77 & $-$0.13 &  0.26 &  0.18 & -22 & 0.13 & 1.87 &  32.8 & 0.85 & 1.6 & -  \\
G018.19$-$00.31 & 18.19 & $-$0.31 &  50.1 &  4.15 & 3.29 & 18.20 & $-$0.44 &  0.39 &  0.27 &  89 & 0.27 & 1.89 &  52.5 & 1.52 & 2.6 & X  \\
G025.64$-$00.11 & 25.64 & $-$0.11 &  93.9 &  2.78 & 3.21 & 25.67 & $-$0.26 &  0.36 &  0.30 &  12 & 0.29 & 1.64 &  24.8 & 1.06 & 1.2 & Y  \\
G024.79+00.09 & 24.79 & 0.09 & 110.4 &  3.25 & 3.17 & 24.62 & 0.17 &  0.48 &  0.37 &   4 & 0.48 & 1.50 &  50.3 & 1.79 & 2.5 & -  \\

\hline
\end{tabular}
\end{sidewaystable}
\clearpage
\begin{sidewaystable}
\caption{\label{clump-list} Properties of the molecular clumps identified in the GRS.}
\scriptsize
\begin{tabular}{cccccccccccccc}\hline\hline
Cloud & Clump &\multicolumn{2}{c}{Peak} & \multicolumn{1}{c}{V$_{LSR}$} & $\Delta \ell$ & $\Delta b$ &  $\Delta$V & \tmb & A &  I$_{peak}$ & I$_{total}$ & N(H$_{2}$) & Flag\\
GRSMC & & $\ell$ & \multicolumn{1}{c}{$b$} \\
\tiny  
& & (\degree) & (\degree) & \multicolumn{1}{c}{(\kms)} & (\degree) & (\degree) & (\kms) & (K) & (deg$^{2}$) & (\Kkms) & (\Kkms\,deg$^{2}$) & ($\times$10$^{21}$ cm$^{-2}$) &  \\
\scriptsize
(1) & (2) & \multicolumn{1}{c}{(3)} & \multicolumn{1}{c}{(4)} & (5) & (6) & (7) & \multicolumn{1}{c}{(8)} & (9) & (10) & \multicolumn{1}{c}{(11)} & (12) & (13) & (14) \\
\hline
G053.59+00.04 &   c1 & 53.57 & 0.06 &  23.5 &  0.08 &  0.08 &  1.95 & 21.52 &  2.00E-02 &   35.65 &  2.39E-01 & 10.6 & V  \\
G053.59+00.04 &   c2 & 53.62 & 0.03 &  23.5 &  0.21 &  0.09 &  1.49 & 20.23 &  4.55E-02 &   34.02 &  2.93E-01 &  9.9 & -  \\
G053.59+00.04 &   c3 & 53.40 & 0.03 &  22.5 &  0.14 &  0.08 &  2.39 &  7.60 &  1.34E-02 &    9.82 &  4.43E-02 &  3.7 & X  \\
G053.59+00.04 &   c4 & 53.49 & $-$0.01 &  21.6 &  0.04 &  0.07 &  1.52 &  5.98 &  4.84E-03 &    3.91 &  8.80E-03 &  2.9 & V  \\
G029.89$-$00.06 &   c1 & 29.87 & $-$0.04 & 100.9 &  0.07 &  0.09 &  2.52 & 16.06 &  2.20E-02 &   46.52 &  3.41E-01 &  7.9 & -  \\
G029.89$-$00.06 &   c2 & 29.96 & $-$0.02 &  97.7 &  0.06 &  0.10 &  2.20 & 15.75 &  1.90E-02 &   47.69 &  1.91E-01 &  7.7 & -  \\
G029.89$-$00.06 &   c3 & 29.91 & $-$0.04 &  98.3 &  0.08 &  0.07 &  2.43 & 14.52 &  1.73E-02 &   46.81 &  2.02E-01 &  7.1 & -  \\
G029.89$-$00.06 &   c4 & 29.91 & $-$0.05 & 101.5 &  0.08 &  0.10 &  2.71 & 14.21 &  2.61E-02 &   44.25 &  3.52E-01 &  7.0 & -  \\
G029.89$-$00.06 &   c5 & 29.93 & $-$0.02 &  98.1 &  0.04 &  0.04 &  2.95 & 13.06 &  5.90E-03 &   36.76 &  6.97E-02 &  6.4 & -  \\
G029.89$-$00.06 &   c6 & 30.01 & $-$0.04 &  93.9 &  0.05 &  0.06 &  2.80 & 12.06 &  1.30E-02 &   27.99 &  1.24E-01 &  5.9 & V  \\
G029.89$-$00.06 &   c7 & 30.00 & 0.01 & 100.2 &  0.08 &  0.08 &  3.04 & 10.27 &  2.07E-02 &   26.91 &  1.58E-01 &  5.0 & -  \\
G029.89$-$00.06 &   c8 & 29.78 & $-$0.26 &  99.8 &  0.11 &  0.08 &  2.02 &  9.42 &  2.24E-02 &   19.33 &  1.25E-01 &  4.6 & -  \\
G029.89$-$00.06 &   c9 & 29.69 & $-$0.17 & 100.2 &  0.10 &  0.10 &  2.32 &  8.38 &  1.98E-02 &   19.14 &  1.45E-01 &  4.1 & X  \\
G029.89$-$00.06 &  c10 & 30.03 & $-$0.51 & 100.5 &  0.13 &  0.06 &  0.81 &  7.96 &  9.87E-03 &    4.14 &  1.61E-02 &  3.9 & -  \\
G029.89$-$00.06 &  c11 & 30.10 & $-$0.26 &  96.0 &  0.07 &  0.09 &  2.02 &  7.44 &  3.25E-03 &    9.42 &  1.11E-02 &  3.7 & -  \\
G029.89$-$00.06 &  c12 & 30.09 & $-$0.29 &  96.8 &  0.03 &  0.05 &  1.62 &  7.38 &  2.80E-03 &    7.50 &  7.72E-03 &  3.6 & -  \\
G029.89$-$00.06 &  c13 & 30.01 & $-$0.26 & 101.7 &  0.12 &  0.17 &  2.45 &  7.21 &  2.68E-02 &   13.08 &  9.75E-02 &  3.5 & -  \\
G029.89$-$00.06 &  c14 & 29.69 & $-$0.22 &  95.4 &  0.07 &  0.06 &  2.79 &  7.21 &  2.31E-03 &    9.57 &  6.18E-03 &  3.5 & X  \\
G029.89$-$00.06 &  c15 & 30.09 & 0.13 &  97.1 &  0.10 &  0.08 &  1.75 &  6.83 &  9.12E-03 &    7.45 &  2.38E-02 &  3.4 & Y  \\
G029.89$-$00.06 &  c16 & 30.31 & $-$0.28 &  99.4 &  0.06 &  0.13 &  0.55 &  6.81 &  2.80E-03 &    4.94 &  6.23E-03 &  3.3 & -  \\
G029.89$-$00.06 &  c17 & 30.26 & $-$0.29 & 100.9 &  0.06 &  0.04 &  0.81 &  6.65 &  2.57E-03 &    4.88 &  4.70E-03 &  3.3 & -  \\
G029.89$-$00.06 &  c18 & 30.04 & 0.01 & 104.5 &  0.09 &  0.08 &  2.59 &  6.31 &  9.38E-03 &    4.68 &  1.22E-02 &  3.1 & V  \\
G029.89$-$00.06 &  c19 & 30.04 & 0.04 & 105.1 &  0.11 &  0.05 &  2.00 &  6.29 &  2.12E-03 &    3.10 &  2.74E-03 &  3.1 & V  \\
G029.89$-$00.06 &  c20 & 30.13 & $-$0.33 &  99.8 &  0.09 &  0.10 &  1.40 &  5.62 &  6.24E-03 &    6.51 &  1.21E-02 &  2.8 & -  \\

\hline
\end{tabular}
\end{sidewaystable}
\clearpage
\begin{sidewaystable}
\scriptsize
\caption{\label{grs-clouds}Summary of the derived properties of the GRS molecular clouds and clumps.}
\centering
\begin{tabular}{lccccccccccccccc}
\tableline
\tableline
\multicolumn{1}{c}{Properties}      & \multicolumn{6}{c}{Clouds}                                    &             & & \multicolumn{6}{c}{Clumps}                                  \\
\cline{2-8} \cline{10-15}
                                    & Min & Max & Mean & Median & Std dev & Slope                   &  K-S test    &&  Min & Max & Mean & Median & Std dev & Slope                \\
\tableline           										                
Peak \tmb\, temperature (K)         & 0.8   &   5.8  &   1.6  &   1.4  &   0.7  &  $-$3.1 $\pm$ 0.2 &0.000000 &&  1.2  &  27.3  &   5.2  &   4.7  &   2.0  &  $-$4.0 $\pm$ 0.3\\
Linewidth (\kms)                    & 0.6   &   9.8  &   3.6  &   3.4  &   1.4  &  $-$2.9 $\pm$ 0.4 &0.001645 &&  0.2  &   8.3  &   1.4  &   1.2  &   0.8  &  $-$3.1 $\pm$ 0.4\\
Semimajor Axis (degs)                   & 0.06  &   1.16 &   0.41 &   0.41 &   0.15 &  $-$5.3 $\pm$ 0.3 &0.000014 &&  0.01 &   0.25 &   0.03 &   0.02 &   0.02 &  $-$2.3 $\pm$ 0.2\\
Semiminor Axis (degs)                   & 0.05  &   0.53 &   0.23 &   0.22 &   0.09 &  $-$2.7 $\pm$ 0.5 &0.000138 &&  0.01 &   0.24 &   0.03 &   0.02 &   0.02 &  $-$2.0 $\pm$ 0.3\\
Position Angle (degs)               & -90.0 &  90.0  &   1.2  &  -1.0  &  43.5  &                   &0.965160 &&       &        &        &        &        &                  \\
Log[Peak \hh\, column density (\cms)]& 20.7 & 23.0   & 21.8   & 21.8   &  0.3   &  $-$2.3 $\pm$ 0.2 &0.000000 &&  20.8 & 22.1   & 21.4   & 21.4   & 0.2    &  $-$4.1 $\pm$ 0.2\\
Excitation Temperature (K)          &  4.1  &  17.5  &   8.8  &   8.5  &   2.0  &  $-$6.8 $\pm$ 0.6 &0.000001 &&       &        &        &        &        &                  \\
Opacity                             &  0.07 &   0.50 &   0.13 &   0.12 &   0.04 &  $-$3.7 $\pm$ 0.3 &0.000231 &&       &        &        &        &        &                  \\
Radius (pc)                         &  1.6  &  97.5  &  24.1  &  22.6  &  12.5  &  $-$4.0 $\pm$ 0.4 &0.002011 &&   1.2 &  27.3  &   5.2  &   4.7  &   2.0  &  $-$4.1 $\pm$ 0.2\\
Log[LTE Mass (\Msun)]               &  2.2  &   5.7  &   4.5  &   4.6  &   0.6  &  $-$1.8 $\pm$ 0.3 &0.000000 &&   0.2 &   5.3  &   2.9  &   2.9  &   0.7  &  $-$1.2 $\pm$ 0.1\\
Log[Virial Mass (\Msun)]            &  1.9  &   5.7  &   4.5  &   4.5  &   0.6  &  $-$1.5 $\pm$ 0.2 &0.000532 &&   0.3 &   5.2  &   2.7  &   2.7  &   0.7  &  $-$1.2 $\pm$ 0.1\\
Log[Density (\cmc)]                 &  0.6  &   2.8  &   1.7  &   1.7  &   0.4  &  $-$1.4 $\pm$ 0.2 &0.000000 &&   0.8 &   3.6  &   2.5  &   2.6  &   0.4  &  $-$2.9 $\pm$ 0.4\\
\tableline
\end{tabular}
\end{sidewaystable}

\end{document}